\documentclass[a4paper,11pt]{article}
\pdfoutput=1 % if your are submitting a pdflatex (i.e. if you have
\usepackage{jheppub} % for details on the use of the package, please
\usepackage[T1]{fontenc} % if needed
\usepackage[utf8]{inputenc}
\usepackage{amsmath,amssymb,amstext,amsfonts} % AMS' mathematical symbols

\usepackage[dvipsnames]{xcolor}
\usepackage{graphicx}
\usepackage{hyperref}
\usepackage{cancel} %for strike-through font
\usepackage{float}
\usepackage{bbm}
\usepackage{mathtools}

\usepackage{lscape}
\usepackage{siunitx}
\usepackage{subcaption}

\usepackage{cleveref}

\usepackage{slashed}
\usepackage{physics}
\usepackage{nicematrix}

\usepackage{comment}

\makeatletter
\renewcommand*\env@matrix[1][\arraystretch]{%
  \edef\arraystretch{#1}%
  \hskip -\arraycolsep
  \let\@ifnextchar\new@ifnextchar
  \array{*\c@MaxMatrixCols c}}
\makeatother

\makeatletter
\g@addto@macro\bfseries{\boldmath}
\makeatother

\def\beq{\begin{equation}}
\def\eeq{\end{equation}}
\def\bsp#1\esp{\begin{split}#1\end{split}}

\usepackage{tikz}
    \usetikzlibrary{positioning}
    \usetikzlibrary{arrows}
    \usetikzlibrary{shapes}
    \usepackage{pgfplots}
    \usetikzlibrary{calc}
    \usetikzlibrary{decorations.markings}
     \usetikzlibrary{decorations.pathmorphing}
    \tikzset{snake it/.style={decorate, decoration=snake}}
\def\centerarc[#1](#2)(#3:#4:#5) % Syntax: [draw options] (center) (initial angle:final angle:radius)
    { \draw[#1] ($(#2)+({#5*cos(#3)},{#5*sin(#3)})$) arc (#3:#4:#5); }
    \usepackage{booktabs}
\usepackage[compat=1.1.0]{tikz-feynman}

\usepackage{contour}
\usepackage{booktabs}
\usepackage{multirow}

\definecolor{cnblue}{RGB}{7,82,154}

\title{\boldmath 
The Four-Loop Ladder Integral Family for the QED Electron Self-Energy}

\author[a]{Felix Forner,} 
\author[b]{Andreas von Manteuffel,}
\author[c]{Christoph Nega,} % \note{Corresponding author.}}
\author[a]{and Lorenzo Tancredi} % \note{Also at Some University.}}

\affiliation[a]{Technical University of Munich, TUM School of Natural Sciences, Physics Department, James-Franck-Straße 1, 85748 Garching, Germany}
\affiliation[b]{Institut für Theoretische Physik, Universität Regensburg, 93040 Regensburg, Germany}
\affiliation[c]{Max Planck Institute for Gravitational Physics (Albert Einstein Institute), Am Mühlenberg 1, 14476 Potsdam, Germany}

\emailAdd{felix.forner@tum.de}
\emailAdd{manteuffel@ur.de}
\emailAdd{christoph.nega@aei.mpg.de}
\emailAdd{lorenzo.tancredi@tum.de}

\preprint{\begin{minipage}[t]{8cm}\begin{flushright} 
TUM-HEP-1622/26
      \end{flushright}\end{minipage}}

\abstract{

We compute the complete set of master integrals of the four-loop ladder integral family contributing to the electron self-energy in Quantum Electrodynamics. We obtain the results through canonical differential equations and write them as iterated integrals over integration kernels related to different parametrizations of the sunrise elliptic curve and a Calabi-Yau threefold. 
The rich structure of geometries and couplings between them requires the introduction of a large number of extra transcendental functions to reach $\epsilon$-factorization. We discuss explicitly how to find a canonical basis for a selection of particularly interesting sectors. Additionally, we compute local series expansions and numerical boundary constants at $p^2=\infty$ and $p^2=m^2$ up to weight seven, and for the polylogarithmic constants we find analytical representations by numerical matching.
The on-shell boundary values of this single family already span all non-polylogarithmic constants entering the planar contribution to the four-loop result for the anomalous magnetic moment $(g-2)$ of the electron.
We obtain precise numerical results across all values of $p^2$ using numerical transport.

}

\begin{document} 
\maketitle
\flushbottom

\section{Introduction}\label{sec:intro}

Quantum Electrodynamics (QED) is at the basis of our most fundamental understanding of the interactions between light and matter. Historically, the study of QED, as the first realistic quantum field theory (QFT), also provided a testing ground for laying 
the foundation for the theory of strong interactions and, eventually, 
the Standard Model of Particle Physics. Among the fundamental objects of study in QED 
are the electron and photon self-energies. Not only are they essential to its UV renormalization,
but they also provide an interesting laboratory to explore the structure of higher-order 
 corrections in perturbative QFT.
For this reason, QED self-energies have attracted considerable interest, 
dating back a long time. In fact, their one-loop calculation is by now textbook material, and 
the first attempt to explore two-loop corrections goes back to more than 70 years ago~\cite{Kallen:1955fb, Sabry}.

In this context, 
the two-loop electron self-energy is of particular importance. Its calculation marked the first appearance in perturbative QFT 
of functions beyond polylogarithms, associated with a family of elliptic curves. 
Owing to these new mathematical structures, after the pioneering calculation by Sabry~\cite{Sabry}, 
this calculation has been revisited several times~\cite{Broadhurst:1993mw,Fleischer:1998dw,Laporta:2004rb}, culminating in a modern evaluation in terms of iterated integrals of modular forms~\cite{Honemann:2018mrb}. 
Since then, many more examples have been identified, 
where the appearing functions can be understood as iterated integrals defined on elliptic curves and more general geometries, see refs.~\cite{Bourjaily:2022bwx,Weinzierl:2022eaz} for reviews and, e.g., refs.~\cite{Aglietti:2007as, Czakon:2008ii,Bonciani:2016qxi, vonManteuffel:2017hms, Blumlein:2018aeq, Adams:2018kez, Adams:2018bsn, Abreu:2019fgk, Abreu:2022cco, Delto:2023kqv, Becchetti:2023yat, Stawinski:2023qtw, Ahmed:2024tsg, Ablinger:2025nnq, Becchetti:2025rrz, Coro:2025vgn, Ahmed:2025osb, Badger:2025ljy, Becchetti:2025qlu, Lellouch:2025rnz, Marzucca:2025eak, Lellouch:2026xdr, Gehrmann:2026qbl, Buccioni:2026zva, Bern:2025wyd, Klemm:2024wtd, Driesse:2024feo, Driesse:2026qiz, Bern:2026oqp} for a (non-exhaustive) selection of examples. Among those are the three-loop contributions to the QED self-energies, whose calculation has been completed only recently \cite{Duhr:2024bzt,Forner:2024ojj}. 

For a long time, problems involving functions beyond polylogarithms 
have escaped the reach of analytic methods. More recently, modern advances on the interplay of 
multi-loop Feynman integrals and algebraic geometry have provided 
novel methods with the potential to overcome these limitations, stimulating a revived interest in these problems.
A crucial development in this direction has been 
the method of differential equations \cite{Kotikov:1990kg, Remiddi:1997ny, Gehrmann:1999as}, improved by the 
choice of a so-called canonical basis~\cite{Henn:2013pwa}. One of the central features of these bases is that they 
satisfy $\epsilon$-factorized systems of differential equations in the kinematic invariants, where $\epsilon$ is the dimensional regulator. In the polylogarithmic case, it is well-known that canonical bases consist of pure integrals with uniform transcendental weight and can be identified by searching for integrands with unit leading singularities~\cite{ArkaniHamed:2010gh, Henn:2020lye}. Several generalizations of canonical bases beyond polylogarithms have been proposed in the literature~\cite{Adams:2016xah, Frellesvig:2021hkr, Pogel:2022ken, Dlapa:2022wdu, Gorges:2023zgv, Duhr:2025lbz, Chen:2025hzq, e-collaboration:2025frv, Bree:2025tug, Duhr:2025xyy, Yang:2025ofz}, as well as approaches that avoid $\epsilon$-factorization~\cite{Chaubey:2025adn,Chaubey:2026egr}.
The construction of these bases can be easily streamlined by the use of AI Agents, as also argued recently in~\cite{PetitRosas:2026qkw}.
In particular, the method proposed in~\cite{Gorges:2023zgv, Duhr:2025lbz} is based on integrand analysis to select suitable candidates, which are then rotated to a fully canonical basis by leveraging the concept of unipotency.
Unipotent functions satisfy differential equations without a homogeneous term and
are at the basis of the definition of multiple polylogarithms on arbitrary geometries~\cite{BrownLevin, Broedel:2017kkb, Broedel:2018qkq}. 
Based on this construction, it was recently argued that, also beyond the polylogarithmic case, canonical integrals can be understood as integrals with constant leading singularities~\cite{Forner:2026vby}. 

The calculation of self-energies in QED at the four-loop order provides an extraordinary testing ground for these new techniques.
In fact, at any number of loops $\ell$, the self-energies are expected to involve Feynman integrals
associated with  Calabi-Yau geometries at least through contributions of so-called banana graphs.
In QED with a single electron, propagators can either be massless or carry the electron mass. 
Thus, at $\ell$ loops, 
we expect to encounter banana integrals with $n \leq \ell+1 $ equal-mass lines and some number of additional massless lines. 
In turn, these are known to be related to $(n-2)$ complex-dimensional Calabi-Yau manifolds~\cite{Vanhove:2014wqa, Vanhove:2018mto, Klemm:2019dbm, Bonisch:2020qmm, Bonisch:2021yfw}, whose properties are very well understood (see also refs.~\cite{Groote:2005ay, Pogel:2022vat, Kreimer:2022fxm} for other studies of banana integrals at any loop order).
In particular, elliptic curves, which correspond to one-dimensional Calabi-Yau manifolds, appear for the first time at two loops~\cite{Bauberger:1994by, Bauberger:1994hx, Bauberger:1994nk, Caffo:1998du, 
Laporta:2004rb, Kniehl:2005bc, Bloch:2013tra, Adams:2013nia, Adams:2014vja, Adams:2015gva, Adams:2016xah, Remiddi:2016gno, Bloch:2016izu},
while at three loops a specific two-dimensional Calabi-Yau geometry appears: a one-parameter K3 surface, which can
be shown to be the symmetric square of the same elliptic curve appearing at two loops~\cite{Joyce1, Primo:2017ipr, Bloch:2014qca, Broedel:2019kmn, Broedel:2021zij, Pogel:2022yat}. From this point of view, then,
the two- and three-loop cases are very special, as they are ultimately related to the same geometry.
On the other hand, starting at four loops, one encounters the first instance of a full-fledged three-dimensional Calabi-Yau manifold.

It follows from fermion number conservation that only bananas with odd (even) numbers of massive lines can appear in the electron (photon) self-energy, 
such that bananas can only contribute with odd (even) dimensional Calabi-Yau geometries. 
This further implies that choosing either the electron or photon self-energy, 
one can only find banana integrals with a new geometry every two loop orders. 
Of course,  geometries beyond polylogarithms can also appear from Feynman integrals 
which are topologically distinct from the $\ell$-loop banana graph.
It is therefore an interesting question whether contributions from other graphs, not related to the banana graph,
can break the pattern identified above at some point in the loop expansion.
In fact, already at three loops, more sectors associated with non-trivial geometries appear~\cite{Duhr:2024bzt}.
Nevertheless, all results available so far~\cite{Duhr:2024bzt, Forner:2024ojj} follow the pattern described above: 
the electron self-energy is polylogarithmic at one loop and elliptic at two and three loops, while the photon self-energy is polylogarithmic
at one and two loops and it involves a one-parameter K3 surface at three loops.
At four loops, no complete results are available yet, but as we will discuss in the following, functions associated with an elliptic curve and a Calabi-Yau threefold appear in the electron self-energy.

\begin{figure}[t]
    \centering
    \includegraphics[width=0.4\linewidth]{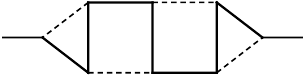}
    \caption{Top sector of the scalar ladder integral family we consider in this work. Massive and massless propagators are depicted by solid and dashed lines, respectively. 
    }
    \label{fig:topsecFe1}
\end{figure}

Motivated by the discussion above, we address in this work the calculation of the ladder integral family contributing to the electron self-energy in QED at four loops shown in~\cref{fig:topsecFe1}.
We choose this particular family because, while being planar and computationally less involved than other families, 
it contains banana subsectors with three and five masses, which suggests a non-trivial interplay between 
a Calabi-Yau threefold and an elliptic curve. 
From the point of view of its mathematical complexity, this makes it an ideal starting point for the computation of the QED self-energies at four loops.
The computation of the electron self-energy is also closely related to one of the most famous quantities in QED, which is the anomalous magnetic moment $(g-2)$ of the electron. The vertex diagrams contributing to $(g-2)$ are obtained from the self-energy diagrams by attaching an external photon to an internal electron line. After expanding to linear order in the photon momentum, as required to extract the magnetic form factor in the on-shell limit for the photon, the resulting integrals involve the same propagators as the self-energy diagrams, with raised propagator powers and additional numerator factors. They can therefore be reduced to the same set of master integrals. Consequently, the on-shell limit of our results is expected to produce the same types of transcendental numbers as those encountered in the four-loop QED corrections to $(g-2)$~\cite{Laporta:1996mq, Laporta:2008sx, Schnetz:2016fhy, Schnetz:2017bko, Laporta:2017okg}. 
In fact, although many planar integral families contribute to the four-loop $(g-2)$, we find that the family in \cref{fig:topsecFe1} already contains the full set of planar non-polylogarithmic constants through weight seven in its on-shell limit.

The remainder of this work is structured as follows. In \cref{sec:general}, we define the integral family under study, outline the method we use to construct a canonical basis, and give an overview of the non-trivial geometries that appear in our computation. 
Then, in \cref{sec:can basis}, we describe how we constructed a canonical basis for all master integrals, highlighting the different mechanisms that require the introduction of new transcendental functions by considering four sectors explicitly: the four-loop equal-mass banana graph (\cref{sec:banana}), two different types of eyeball graphs (\cref{sec:eyeball,sec:eyeball2}), and the cateye graph (\cref{sec:cateye}). 
Our results are given in \cref{sec:results}, where we discuss both the general analytic solution in terms of iterated integrals in \cref{sec:analyticresults}, as well as series solutions and boundary constants, both at large momentum and at the on-shell point, in \cref{sec:series and boundaries}. 
At the on-shell point, we recover the classes of transcendental numbers known from the four-loop $(g-2)$ of the electron. 
We also comment on the numerical evaluation of our results in \cref{sec:numerical} and we conclude in \cref{sec:conclusion}.
 \Cref{sec:allgraphs,app:precan,app:Gfunctions} contain graphs for our basis integrals as well as our pre-canonical basis together with the definitions of the special functions needed to obtain the canonical basis.
We provide our results for the canonical basis, the differential equations, the boundary conditions, and the series expansion solutions in computer-readable format in a Zenodo repository~\cite{Zenodo}.

\section{The integral family and its geometries}
\label{sec:general}

In this work, we consider the scalar four-loop ladder integral family
\begin{equation}
\mathcal{I}_{n_1 \,\dots\, n_{14}}
= e^{4 \epsilon \gamma} \int \prod_{j=1}^4 \frac{\mathrm d^D k_j}{i \pi^{D/2}} \frac{1}{D_1^{n_1} \cdots D_{14}^{n_{14}}} \, ,
\label{eq:intfam}
\end{equation}
where $\gamma$ is the Euler-Mascheroni constant and the inverse propagators $D_j$ are
\begin{equation}
    \begin{aligned}
        D_{1\phantom{0}}  &= k_2^2\,, &
        D_{2\phantom{0}}  &= \left(p-k_1\right)^2\,, &
        D_{3\phantom{0}}  &= \left(k_1-k_3\right)^2\,, \\
        D_{4\phantom{0}}  &= \left(k_2-k_4+p\right)^2\,, &
        D_{5\phantom{0}}  &= k_1^2-m^2\,, &
        D_{6\phantom{0}}  &= k_3^2-m^2\,, \\
        D_{7\phantom{0}}  &= k_4^2-m^2\,, &
        D_{8\phantom{0}}  &= \left(k_2+p\right)^2-m^2\,, &
        D_{9\phantom{0}}  &= \left(k_2-k_4\right)^2-m^2\,, \\
        D_{10} &= \left(-k_1+k_3+p\right)^2-m^2\,, &
        D_{11} &= \left(k_1+k_2-k_3-k_4\right)^2-m^2\,, &
        D_{12} &= k_3\cdot k_4\,, \\
        D_{13} &= k_2\cdot k_3\,, &
        D_{14} &= k_1\cdot k_4\,. & &
    \end{aligned}
\end{equation}
We work in dimensional regularization, such that the space-time dimension $D$ and the dimensional regulator $\epsilon$ are related through $D=4-2\epsilon$. This integral family is of physical interest since it enters the computation of the electron self-energy in QED at four loops.

Following standard conventions, we group the master integrals into sectors, where each sector is labeled by its \textit{sector ID}
\begin{align}\label{eq:sec id}
    \text{ID} = \sum\limits_{j \,\text{with}\,n_j>0} 2^{j-1}\,.
\end{align}
For any sector, we refer to the set of inverse propagators $D_j$ with $n_j>0$ as \textit{active propagators}. 
We then call the sector with the highest number of active propagators the \textit{top sector}. 
For our purposes, the top sector has active propagators $D_1$ to $D_{11}$ and corresponds to the graph shown in \cref{fig:topsecFe1}, which is also the scalar version of one of the Feynman diagrams contributing to the QED electron self-energy. 
Here and in all other scalar Feynman integral graphs, we depict massive propagators with solid lines and massless ones with dashed lines. 
We divide every integral by an appropriate power of $m^2$ to make it dimensionless, such that all quantities depend only on the single dimensionless variable
\begin{equation}
    z=\frac{m^2}{p^2}\,.
\end{equation} 
In practice, this normalization is equivalent to setting $m^2=1$.

We reduce all scalar integrals to a set of independent master integrals using symmetry and integration-by-parts (IBP) relations~\cite{Tkachov:1981wb, Chetyrkin:1981qh, Laporta:2004rb}. 
For this computation, we employ 
\texttt{Reduze 2}~\cite{Studerus:2009ye, vonManteuffel:2012np}, 
and \texttt{Finred}, a private implementation of the Laporta algorithm using finite-field sampling and rational reconstruction~\cite{vonManteuffel:2014ixa, Peraro:2016wsq}.
The reduction leads to a total of $160$ master integrals in $72$ sectors, with up to $6$ master integrals per sector. An overview with all sector IDs and the corresponding graphs can be found in \cref{sec:allgraphs}.

Our goal here is to compute these master integrals with the method of differential equations~\cite{Kotikov:1990kg, Remiddi:1997ny, Gehrmann:1999as}. 
Collecting all masters in a vector $\vec I$, differentiating with respect to $z$ and reducing the result back to master integrals yields a linear system
\begin{equation}
  \partial_z \vec I(z,\epsilon) = A(z,\epsilon)\,\vec I(z,\epsilon)\,.
\end{equation}
When the master integrals are ordered by sectors and start with the tadpole integral, the matrix $A$ is block lower-triangular. 
This system can be solved most efficiently if one finds a \textit{canonical basis}~\cite{Henn:2013pwa} 
$\vec J = T(z,\epsilon)\,\vec I$, 
for which the dependence on $\epsilon$ factorizes
\begin{equation}\label{eq:candeq}
  \partial_z \vec J(z,\epsilon) = \epsilon\, B(z)\,\vec J(z,\epsilon)\,.
\end{equation}
The solution of \cref{eq:candeq} can then be written order by order in $\epsilon$ in terms of Chen iterated integrals~\cite{ChenSymbol} over the entries of $B(z)\mathrm{d}z$.
Thus, the analytic properties of the solution are encoded in the matrix $B(z)$.
If all entries in $B(z)\mathrm{d}z$ are logarithmic forms, canonical bases are by now routinely constructed by selecting integrands in $\mathrm{d}\!\log$-form and with constant leading singularities~\cite{ArkaniHamed:2010gh, Henn:2013pwa, Henn:2020lye}. 
This is not the case for the family considered here, as new differential forms associated with non-trivial geometries, including elliptic curves and Calabi-Yau manifolds, appear.

Various methods have been developed to construct a canonical basis in these more general cases.
Here, we follow the approach of~\cite{Gorges:2023zgv, Duhr:2025lbz}, in the formulation in terms of integrand analysis and leading singularities put forward in~\cite{Forner:2026vby}. 
Our construction proceeds bottom-up and sector by sector.
For each individual sector, we have three main steps:
\begin{enumerate}
  \item For a given sector, we first identify the geometry associated with its maximal cut~\cite{Primo:2016ebd, Primo:2017ipr},
  and choose a starting basis whose integrands correspond to natural differential forms on this geometry.
  \item We then normalize the leading singularities of these integrals on their maximal cut at $\epsilon=0$. Beyond polylogarithms, this amounts to dividing by periods of the underlying geometry and their derivatives. Equivalently, the rotation is given by the inverse of the semi-simple part of the period matrix (also known as Wronskian)~\cite{Gorges:2023zgv, Duhr:2025lbz, Forner:2026vby}.
  
  \item The resulting differential equations are, in general, not yet $\epsilon$-factorized, both within a sector and in the couplings between different sectors. The remaining terms are removed by a last rotation, which requires introducing new transcendental functions, defined as (iterated) integrals over periods and rational functions. As shown in~\cite{Forner:2026vby}, these functions correspond to further leading singularities of the corresponding integrals, either at higher orders in $\epsilon$ or on non-maximal cuts.
\end{enumerate}

We stress here that this split is not strictly necessary.
One could alternatively proceed by attempting a complete integrand analysis of all integrals involved on all cuts as described in~\cite{Forner:2026vby}, and this is expected to provide all information necessary to build a canonical basis in one step. 
On the other hand, we find it much more practical to proceed block by block. 
Compared to a global approach, this allows us to isolate the different geometries and also to preserve the block triangular form of the original differential equations.

We start by determining the geometry of all 
minimally coupled blocks in this problem.
For this, we find especially convenient the (loop-by-loop) Baikov representation~\cite{Baikov:1996iu, Frellesvig:2017aai, Frellesvig:2024ymq}, which we generate with \texttt{SOFIA}~\cite{Correia:2025wtb}.
Since \texttt{SOFIA} chooses its own momentum routing, the integrals we construct are often defined in other families, which we then map to our family using \texttt{Reduze\;2}. 
Depending on the sector, the analysis is simplest close to $D=2$ or $D=4$ dimensions.
Integrals found in $D=2-2\epsilon$ dimensions are expressed in terms of four-dimensional integrals using dimensional shift relations~\cite{Tarasov:1996br, Lee:2009dh}. 
In this way, we find that out of the total of $72$ sectors, $N_{\rm pol}=53$ sectors are polylogarithmic on their maximal cut, while the remaining ones are associated with only two different types of geometries.
In detail, $N_{\rm ell}=18$ sectors involve a family of elliptic curves, and a single sector, the equal-mass four-loop banana graph, involves a family of Calabi-Yau threefolds. 
This is precisely what one expects from the banana graphs with three and five massive propagators contributing to our family.
In particular, although several of the elliptic sectors are not banana graphs, and in one case the elliptic curve appears in a different parametrization, we do not find any geometry beyond those of the banana graphs.
Nevertheless, the full complexity of the problem is determined by 
how these geometries couple. 
The Calabi-Yau threefold enters sectors whose maximal cuts are polylogarithmic or 
elliptic through their coupling to the five-mass banana. In the cat-eye sector, two distinct elliptic curves, obtained by evaluating the sunrise family at different kinematic arguments, already couple on the maximal cut.

In the remainder of this section, we describe the main properties of the various geometries we encounter.
We will focus on the behavior of their periods locally close to two representative points, $z=0$ and $z=1$. 
While $z=0$ represents a point of maximal unipotent monodromy (MUM) for all geometries involved, $z=1$ is the on-shell limit and thus directly relevant for the $(g-2)$ of the electron.

\subsection{The family of elliptic curves}
\label{sec:elliptic}

\begin{figure}[t]
    \centering
    \includegraphics[width=\linewidth]{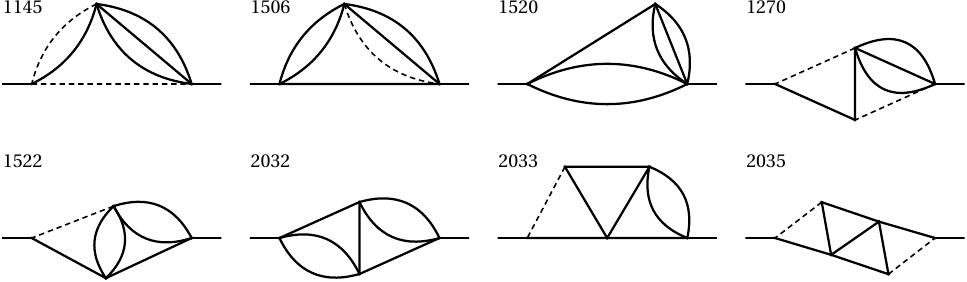}
    \caption{Graphs of all sectors with an elliptic curve on the maximal cut, excluding graphs with factorized banana subgraphs and the cat-eye graph, which is shown separately in \cref{fig:cateye}.}
    \label{fig:ellipticMCs}
\end{figure}

\begin{figure}[t]
    \centering
    \includegraphics[width=0.35\linewidth]{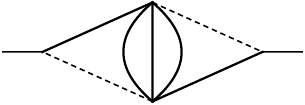}
    \caption{The cat-eye graph, corresponding to sector 1267.}
    \label{fig:cateye}
\end{figure}

The elliptic sectors in our family are of three types. 
The first involves banana graphs or banana subgraphs with three massive propagators. 
Among these are the four-loop banana graph with two massless propagators, as well as many factorized graphs containing either the two-loop equal-mass sunrise graph or the three-loop banana graph with one massless propagator as a subgraph. 
Second, interestingly we encounter eight graphs which are not of banana type, 
shown in \cref{fig:ellipticMCs}, but whose maximal cuts reveal the same geometry as the sunrise graph. 
Finally, we find that the cat-eye graph shown in \cref{fig:cateye} is elliptic on the maximal cut, 
but involves two elliptic curves at the same time.

The first two types are associated with the geometry of the equal-mass sunrise graph, 
which has been studied extensively in the past~\cite{Sabry, Broadhurst:1993mw, Laporta:2004rb, MullerStach:2011ru, Bloch:2013tra, Remiddi:2013joa}, 
possibly with extra punctures~\cite{Bonisch:2021yfw}. 
A summary of all aspects relevant for our purpose, including local series expansions around all singular points, 
may be found in~\cite{Duhr:2024bzt}. 
Here, we recall that the maximal cut of the sunrise graph is related to the period integral $\int \mathrm{d}x/y$ over the family of elliptic curves
\begin{equation}
    y^2 = P_4(x;1/z) = x\big(x-4\big) Q(x,1/z)\,, \quad \text{where}\quad Q(a,b)= 1 + a^2 + b^2 - 2 a b -2a -2b\,. \label{eq:sunrisecurve}
\end{equation}
The periods of this geometry satisfy the second-order Picard-Fuchs equation
\begin{equation}
\mathcal{L}_2\varpi_{i}(z)=\left[
\frac{\partial^2}{\partial z^2}
+\frac{(3 z-1) (3 z+1)}{(z-1)  (9 z-1)z}%\dv{z}
\frac{\partial}{\partial z}-\frac{(3 z-1)}{(z-1) (9 z-1)z^2}
\right] \varpi_{i}(z) = 0\,,\qquad i=0,1\,. \label{eq:deqsunperiods}
\end{equation}
In the rotation to the canonical basis, only the holomorphic period $\varpi_0$ and its first derivative enter, 
together with the determinant of the period matrix of $\mathcal{L}_2$, which is rational.
We choose and normalize the periods $\varpi_0$, $\varpi_1$ as in~\cite{Duhr:2024bzt}, such that $\varpi_0$ is always holomorphic close to a given singular point of the elliptic curve.

The third type is given by the \textit{cat-eye graph} in \cref{fig:cateye}, which has five master integrals. On its maximal cut, we find two families of elliptic curves defined through
\begin{align}\label{eq:defellcurvecateye}
    y^2 = P_4(x;2-1/z) \qquad \text{and} \qquad y^2 = P_4\left(x;z\right)\,,
\end{align}
with $P_4$ as given in \cref{eq:sunrisecurve}. We derive these equations explicitly from the Baikov representation in \cref{sec:cateye}.
Both are given by the same polynomial as the sunrise curve $y^2 = P_4(x;1/z)$, evaluated at a different argument. Hence, all three equations define the same family of elliptic curves in different parametrizations, and no new geometry appears.

Of course, while all elliptic curves are related to the one of the equal-mass sunrise graph,  
fixing a specific value of $z$, the three corresponding elliptic curves are different, as well as their periods and so in practice we have to consider them as independent different elliptic curves in what follows. 
We define the periods $\varpi^{\text{(shift)}}_{0}(z)$, $\varpi^{\text{(shift)}}_{1}(z)$ as the solutions of \cref{eq:deqsunperiods} after shifting $1/z\to2-1/z$ and, analogously, the periods $\varpi^{\text{(inv)}}_{0}$, $\varpi^{\text{(inv)}}_{1}$ as the same solution after transforming $1/z\to z$. 
Also here, locally close to a given MUM point, 
we choose $\varpi^{\text{(shift)}}_{0}$ and $\varpi^{\text{(inv)}}_{0}$ to be holomorphic. Interestingly, 
the five master integrals of the cat-eye graph naturally split into groups of $2+2+1$, unveiling a structure closely related to the ice cone family studied in~\cite{Duhr:2022dxb}. We discuss how to choose a good starting basis for this sector in \cref{sec:cateye}.

As anticipated above, we will consider explicitly the expansion points $z=0$ and $z=1$, which correspond physically to the large momentum ($m^2/p^2\to 0$) and on-shell ($p^2\to m^2$) limits, respectively.

At $z=0$, our choices of elliptic periods admit the local series expansions
\begin{equation}
\begin{aligned}
  \varpi_0^{[0]}(z) &=  z+3 z^2+15 z^3+93 z^4 + \mathcal O(z^5) \,, \\
  \varpi_1^{[0]}(z) &= \varpi_0^{[0]}(z)\log(z)  +4 z^2+26 z^3+\frac{526 z^4}{3} + \mathcal O(z^5)\,, \\
  \varpi_0^{\text{(shift)}[0]}(z)&=  z - z^2 + 7 z^3 - 31 z^4 + \mathcal O(z^5)\,, \\
  \varpi_1^{\text{(shift)}[0]}(z)&= \varpi_0^{\text{(shift)}[0]}(z)\log(z) -2 z^2 +10 z^3 -\frac{158 z^4}{3} + \mathcal O(z^5)\,, \\
  \varpi_0^{\text{(inv)}[0]}(z)&= 1 +\frac{z}{3} +\frac{5 z^2}{27} +\frac{31 z^3}{243} + \frac{71 z^4}{729} + \mathcal O(z^5)\,,  \\
  \varpi_1^{\text{(inv)}[0]}(z)&= \varpi_0^{\text{(inv)}[0]}(z) \log(z) +\frac{4 z}{9} +\frac{26 z^2}{81} +\frac{526 z^3}{2187} +\frac{1253 z^4}{6561} + \mathcal O(z^5)\,.
  \label{eq:ellipticperiodexpansions infty}
\end{aligned}
\end{equation}

Here, and in the following, we write $f^{[z_0]}(z)$ for the series expansion of a function $f(z)$ around $z=z_0$. For the point $z=1$, we transform to the local variable $u=1/z-1$ and expand around $u=0$. Then, in the same fashion as before, our periods are defined by
\begin{equation}
\begin{aligned}
  \varpi_0^{[1]}(u) &= 1 -\frac{u}{4} +\frac{5 u^2}{32} -\frac{7 u^3}{64} + \frac{173 u^4}{2048}+ \mathcal O(u^5) \,, \\
  \varpi_1^{[1]}(u) &= \varpi_0^{[1]}(u)\log(u) -\frac{3 u}{8} +\frac{33 u^2}{128}-\frac{25 u^3}{128} +\frac{2561 u^4}{16384}+ \mathcal O(u^5)\,, \\
  \varpi_0^{\text{(shift)}[1]}(u)&= 1 +\frac{u}{4} +\frac{5 u^2}{32} +\frac{7 u^3}{64} +\frac{173 u^4}{2048}+ \mathcal O(u^5)\,, \\
  \varpi_1^{\text{(shift)}[1]}(u)&= \varpi_0^{\text{(shift)}[1]}(u) \log(u) + \frac{3 u}{8} +\frac{33 u^2}{128} +\frac{25 u^3}{128} + \frac{2561 u^4}{16384}+\mathcal O(u^5)\,, \\
  \varpi_0^{\text{(inv)}[1]}(u)&=1 +\frac{u}{4} -\frac{3 u^2}{32} +\frac{3 u^3}{64} -\frac{51 u^4}{2048}+ \mathcal O(u^5)\,,  \\
  \varpi_1^{\text{(inv)}[1]}(u)&= \varpi_0^{\text{(inv)}[1]}(u) \log(u) -\frac{5 u}{8} +\frac{17 u^2}{128} -\frac{23 u^3}{384} +\frac{2051 u^4}{49152}+ \mathcal O(u^5)\,.
  \label{eq:ellipticperiodexpansions 1}
\end{aligned}
\end{equation}

\subsection{The family of Calabi-Yau threefolds}
\label{sec:CY3}
 
\begin{figure}[t]
    \centering
    \includegraphics[width=0.3\linewidth]{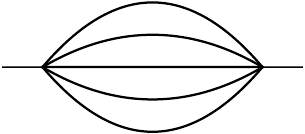}
    \caption{The four-loop equal-mass banana graph, corresponding to sector 1264.}
    \label{fig:banana}
\end{figure}

The only sector with a geometry beyond elliptic curves on its maximal cut is the equal-mass four-loop banana graph, shown in \cref{fig:banana}. 
As is well known~\cite{Vanhove:2018mto, Bonisch:2020qmm, Bonisch:2021yfw}, its maximal cut involves 
the periods of a one-parameter family of Calabi-Yau threefolds. This is easy to see by analyzing the integrand of the maximal cut in the loop-by-loop Baikov representation, which reveals the defining polynomial of the Calabi-Yau threefold
\begin{align}
   y^2=  P_{4,4,4}(x_1,x_2,x_3;1/z) = x_1 (x_1-4) Q(x_1,x_2) Q(x_2, x_3) Q(x_3, 1/z)\,, 
\label{eq:polyCY3}
\end{align}
with $Q$ as defined in \cref{eq:sunrisecurve}. 
Even though this geometry appears on the maximal cut of a single sector, it has the largest impact on the structure of our canonical basis, since the same square root appears, with additional poles, in the integrands of several higher sectors (see \cref{sec:eyeball,sec:eyeball2}) beyond their maximal cut. 
Therefore, $\epsilon$-factorizing the full family requires introducing many functions built from the Calabi-Yau periods.

Let us focus on the properties of the Calabi-Yau geometry. 
It introduces four periods $\pi_i$, $i=0,1,2,3$, defined as the solutions of the fourth-order Picard-Fuchs equation
\begin{equation}
    \begin{split}
        &\Bigg[ %\dv[4]{z} 
        \frac{\partial^4}{\partial z^4}
        +\frac{2 \left(900 z^3-777 z^2+70 z-1\right)}{(z-1)  (9 z-1) (25 z-1)z}%\dv[3]{z} 
        \frac{\partial^3}{\partial z^3}
        +\frac{\left(3150 z^3-1839 z^2+98 z-1\right)}{(z-1) (9 z-1) (25 z-1)z^2}%\dv[2]{z}
        \frac{\partial^2}{\partial z^2}\\
        &+\frac{(15 z-1) \left(60 z^2-15 z-1\right)}{(z-1) (9 z-1) (25 z-1)z^3}\frac{\partial}{\partial z}+ \frac{(5 z-1) }{(z-1) (9 z-1) (25 z-1)z^4}
    \Bigg]\pi_i(z) = 0\,,\quad i=0,1,2,3\,. \label{eq:deqCY3}
    \end{split}
\end{equation}
We first derive series expansions around the MUM point $z=0$, where the four indicials are all equal to unity, and logarithms up to power three appear. 
We choose the periods as
\begin{equation}
    \begin{aligned}
    \pi_0^{[0]}(z) &= z +5 z^2 +45 z^3 +545 z^4 + \mathcal O(z^5) \,, \\
    \pi_1^{[0]}(z) &= \pi_0^{[0]}(z)\log(z) + \Sigma^{[0]}_1\, , \\
    \Sigma^{[0]}_1(z) &= 8 z^2 +100 z^3 +\frac{4148 z^4}{3} + \mathcal O(z^5) \,, \\
    \pi_2^{[0]}(z) &= \frac{1}{2!}\pi_0^{[0]}(z)\log^2(z) + \Sigma^{[0]}_1(z)\log(z) +\Sigma^{[0]}_2(z) \, , \\
    \Sigma^{[0]}_2(z) &= z^2 +\frac{197 z^3}{4} +\frac{33637 z^4}{36}  + \mathcal O(z^5) \,, \\
    \pi_3^{[0]}(z) &= \frac{1}{3!}\pi_0^{[0]}(z)\log^3(z) +\frac{1}{2!}\Sigma^{[0]}_1(z)\log^2(z) + \Sigma^{[0]}_2(z)\log(z) \\
    &\qquad-2 z^2 -\frac{89 z^3}{4} -\frac{19295 z^4}{108}  + \mathcal O(z^5) \,.
    \end{aligned}\label{eq:CY3periodsz}
\end{equation}
In principle, we could have chosen different linear combinations of~\cref{eq:CY3periodsz} as our periods. 
At the MUM point, however, where the full tower of logarithms appears, \cref{eq:CY3periodsz} is the natural choice 
to define the transcendental weight of the corresponding master integrals consistently with the polylogarithmic case.
In technical terms, the periods in~\cref{eq:CY3periodsz} are graded according to their
monodromy filtration weight.

 With this definition, the period matrix 
$\Pi_{ij} = \partial_z^{i-1} \pi_{j-1}$
for $1\leq i,j \leq 4$, satisfies quadratic relations related to Griffiths transversality~\cite{MR0260733, Bonisch:2021yfw}
\begin{align}  
\label{eq:griffiths}
   0 &= \Pi\cdot \begin{pmatrix}
         0 & 0 & 0 & 1 \\
 0 & 0 & -1 & 0 \\
 0 & 1 & 0 & 0 \\
 -1 & 0 & 0 & 0
    \end{pmatrix} \cdot \Pi^T - \begin{pmatrix}
         0 & a_1 & a_2 & a_3 \\
 -a_1 & 0 & -a_3 & 0 \\
 -a_2 & a_3 & 0 & 0 \\
 -a_3 & 0 & 0 & 0
    \end{pmatrix}^{-1} \, , \nonumber\\
    a_1 &= -\frac{1}{z}+28 -285 z + 450 z^2\,, \qquad\quad \,\,
    a_2 = -1 +70 z -777 z^2+900 z^3\,, \\
    a_3 &= -z +35 z^2-259 z^3  + 225 z^4\,. \nonumber
\end{align}

Furthermore, for our application, not all periods will be relevant. Thanks to the quadratic relations above, 
one can prove that the semi-simple part of the period matrix depends only on two transcendental functions~\cite{Duhr:2025lbz, Forner:2026vby}, which are the holomorphic period $\pi_0$ and the structure series $\alpha$, defined through
\begin{align}\label{eq:def yukawa coupling}
    \alpha(z)^{-1} =  z\,\frac{\partial}{\partial z} \left(\frac{\pi_1(z)}{\pi_0(z)}\right) \,.
\end{align}
For a more thorough definition of the structure series and its relations to the mirror map and $Y$-invariants or Yukawa-couplings of the Calabi-Yau manifold, we refer to~\cite{Duhr:2022dxb, Bogner:2013kvr, Bognerthesis}.
All other functions needed in the rotation to the canonical basis, which we call collectively $G$-functions in the following, are (iterated) integrals over $\pi_0$, $\alpha$, the elliptic periods introduced above, and rational functions, see \cref{sec:can basis}. 
As we will see there, our choice in \cref{eq:CY3periodsz} guarantees that the 
series expansions of all these quantities around $z=0$ are free of logarithms.

The situation is less transparent when expanding around singular points that are not MUM since the full tower of logarithms does not appear. In these cases, we first derive an arbitrary basis of solutions of the differential operator and take four generic linear combinations of these solutions. We then choose the coefficients of these combinations 
such that the series expansions of $\pi_0$, $\alpha$ and the $G$-functions are free of logarithms, 
the differential equation matrix has only simple poles, and the Griffiths relations \cref{eq:griffiths}, 
transformed to the singular point under study, are satisfied. While we always managed to satisfy these conditions in all cases we considered in this and previous calculations, it would be interesting to prove, in general, under which assumptions one should expect this to be possible.

For $z=1$ (corresponding to $p^2=m^2$), we expand around $u=0$ in the local variable $u=1/z-1$, following the
argument above.  
This is a so-called conifold point of the Calabi-Yau geometry with indicials $\{0, 1, 1, 2\}$, and we take the periods
\begin{equation}\label{eq:expansions1}
    \begin{aligned}
    \pi_0^{[1]}(u) &=1 -\frac{u}{6} +\frac{u^3}{108} -\frac{59 u^4}{5184}+ \mathcal O(u^5) \,, \\
    \pi_1^{[1]}(u) &=-u +\frac{u^2}{2} -\frac{11 u^3}{32} + \frac{25 u^4}{96}+ \mathcal O(u^5) \,, \\
    \pi_2^{[1]}(u) &= - \frac{1}{192}\pi_1^{[1]}(u)\log(u) -\frac{u^2}{768} +\frac{89 u^3}{73728} -\frac{115 u^4}{110592} + \mathcal O(u^5) \,, \\
    \pi_3^{[1]}(u) &= -\frac{u^2}{384} +\frac{7 u^3}{3456} -\frac{2197 u^4}{1327104} + \mathcal O(u^5) \,.
    \end{aligned}
\end{equation}
We illustrate how these conditions constrain the choice of the periods with an explicit example in \cref{sec:onshell}, where we use the expansions around $z=1$.

\section{Constructing a Canonical Basis}
\label{sec:can basis}

With the geometries of \cref{sec:general} at hand, we can proceed with steps 2 and 3 of our general procedure to obtain a canonical basis for our family of $160$ master integrals.
In particular, having studied the integrands related to all maximal cuts in the previous section allows us to identify a suitable pre-canonical basis, see \cref{app:precan}, which consists only of rational linear combinations of Feynman integrals, i.e., where we have not introduced any square roots or transcendental functions yet.
From it, we proceed to step 2 and normalize the maximal-cut leading singularities of these integrals at $\epsilon=0$. 
This also corresponds to integrating out the semi-simple part of the period matrix of each coupled block~\cite{Gorges:2023zgv, Duhr:2025lbz, Forner:2026vby}, and the required rotation introduces square roots and the periods of the geometries of \cref{sec:general}. 

As anticipated above, the differential equations obtained in this way are, in general, not yet $\epsilon$-factorized. 
The remaining non-factorized entries are removed by a further rotation, which requires the introduction of new transcendental functions. 
All these functions, as well as the square roots and periods, can be defined as solutions of differential equations with rational coefficients.
Equivalently, they can all be written as (iterated) integrals over rational functions, square roots, and the periods of \cref{sec:general}. 
In total, we need $27$ such functions, $G_1,\dots,G_{27}$, whose defining differential equations are listed in \cref{app:Gfunctions}. 
The full rotation to the canonical basis, the corresponding $\epsilon$-factorized differential equations, and the series expansions of the transcendental functions are provided in the Zenodo repository~\cite{Zenodo}.

Before showing explicitly how to derive some of these functions for some representative cases, it is instructive to sort them according to the mechanism that makes them necessary.
\begin{itemize}

  \item $G_{23}$ is the simplest, as it appears on the maximal cut of an elliptic sector with 6 master integrals, see the graph of sector 1522 in~\cref{fig:ellipticMCs}, and corresponds to a third-kind form on that geometry. This type of $G$-function has been discussed at length in previous publications~\cite{Gorges:2023zgv, Duhr:2025lbz} and we  do not discuss it further here.
  
  \item $G_1,\dots, G_4$ are needed to $\epsilon$-factorize the four-loop equal-mass banana sector, i.e., the only sector with a Calabi-Yau threefold on its maximal cut. Their origins can be traced to new leading singularities arising in the higher-$\epsilon$ orders of the three master integrals associated with differential forms with higher poles.
  
  \item $G_5$ arises from couplings of sectors that are polylogarithmic on their maximal cut, to the Calabi-Yau threefold through their subsectors.
  
  \item $G_6,\dots,G_9$, $G_{19},\dots,G_{22}$ and $G_{24},\dots,G_{27}$ arise from sectors that are elliptic on their maximal cut and couple to the Calabi-Yau threefold.
  
  \item $G_{10},\dots,G_{18}$ arise from the cat-eye graph, which couples two different elliptic curves to each other ($G_{10}$) and to the Calabi-Yau threefold ($G_{11},\dots,G_{18}$).

\end{itemize}
Remarkably, the Calabi-Yau threefold enters the large majority of these functions, although it only appears on the maximal cut of a single sector.

In the following subsections, we discuss one representative sector for each of the last four mechanisms, in order of increasing complexity: the four-loop equal-mass banana graph, where the Calabi-Yau threefold appears on the maximal cut (\cref{sec:banana}); an eyeball graph, which is polylogarithmic on its maximal cut but couples to the Calabi-Yau threefold (\cref{sec:eyeball}); a double-eyeball graph, where an elliptic curve and the Calabi-Yau threefold couple (\cref{sec:eyeball2}); and the cat-eye graph, where two different elliptic curves couple to each other already on the maximal cut and additionally couple to the Calabi-Yau threefold (\cref{sec:cateye}). 
In all examples, as a reference, we work close to the MUM point at $z=0$, which allows us to interpret the local monodromy filtration weight (number of logarithms in the solutions) in terms of the usual polylogarithmic transcendental weight.

\subsection{The four-loop equal-mass banana graph}
\label{sec:banana}
For an example of the second type of $G$-functions, we consider the maximal cut of the four-loop banana graph shown in \cref{fig:banana}, which corresponds to sector 1264 in our integral family defined in \cref{eq:intfam}.
It is easy to show by integrand analysis that $I_\text{Ban4L} = I_{0 0 0 0 1 1 1 1 0 0 1 0 0 0}$ corresponds to the holomorphic first-kind differential form on the Calabi-Yau threefold of \cref{eq:polyCY3}. 
From our starting basis, we transform to the derivative basis $\vec I_d=(I_\text{Ban4L}, \partial_z I_\text{Ban4L}, \partial_z^2 I_\text{Ban4L}, \partial_z^3 I_\text{Ban4L})^T$, which produces integrals that are related to the forms of second kind on the manifold.
The new basis satisfies the system of differential equations
\begin{align}
    \partial_z \vec I_d = \begin{pmatrix}
        0 & 1 & 0 & 0 \\
        0 & 0 & 1 & 0 \\
        0 & 0 & 0 & 1 \\
        c_1 & c_2 & c_3 & c_4
    \end{pmatrix}\cdot
    \vec I_d \,,
\label{eq:PF_system}
\end{align}
where
\footnotesize
\begin{align}
    c_1 &= \frac{(72z+24)\epsilon^4 +(50-42z)\epsilon^3
            +(35-103z)\epsilon^2+(10-42z)\epsilon+1-5z}{(z-1)z^4(9z-1)(25z-1)}\,,\nonumber\\[5pt]
    c_2 &= \frac{(414z^2+84z-50)\epsilon^3-(450z^3+363z^2+35)\epsilon^2-(1350z^3+492z^2-84z+10)\epsilon-900z^3+285z^2-1}{(z-1) z^3 (9 z-1) (25 z-1)} \,,\nonumber\\
    c_3 &= \frac{(-225z^3-363z^2-343z+35)\epsilon^2+(-2025z^3-1554z^2+315z)\epsilon -3150z^3+1839z^2-98z+1}{(z-1) z^2 (9 z-1) (25 z-1)}\,,\nonumber\\
    c_4 &= \frac{(-450z^3-518z^2+210z-10)\epsilon-1800z^3+1554z^2-140z+2}{(z-1) z (9 z-1) (25 z-1)}\,.
\end{align}
\normalsize

The last row in \cref{eq:PF_system} also provides the homogeneous fourth-order differential equation satisfied by $I_\text{Ban4L}$ and, by construction, reduces for $\epsilon=0$ to the Picard-Fuchs equation in \cref{eq:deqCY3}. Its solutions at $\epsilon=0$ are the periods defined through the local behavior given in \cref{eq:CY3periodsz}.

We can now normalize the leading singularities at $\epsilon=0$, using only the two functions $\pi_0$ and $\alpha$ defined in \cref{sec:CY3}. 
Since the basis is spanned by four forms, there are also four cycles with a leading singularity on each. 
Throughout this work, we write $\text{LS}\left[f(x)\right]_C$ to denote the leading singularity of the function $f(x)$ associated with the cycle $C$. 
Here, we write $C_0, C_1, C_2, C_3$ for the four cycles, where $C_0$ is the holomorphic one. 
For the first integral, we have
\begin{align}
    \text{LS}[I_\text{Ban4L}]_{C_0} = \pi_0 + \mathcal{O}(\epsilon)\,.
\end{align}
Thus, we define the normalized integral $J_1 = I_\text{Ban4L}/\pi_0$. 
For the second integral, the derivative of the properly normalized integral $J_1$ is closer to a canonical integral than the derivative of $I_\text{Ban4L}$.
In fact, while its leading singularity on the first cycle, $\text{LS}[\partial_z J_1]_{C_0}$, is zero by construction, one finds on the second cycle
\begin{align}
    \text{LS}[\partial_z J_1]_{C_1} = \frac{\partial}{\partial z} \left(\frac{\pi_1}{\pi_0}\right) + \mathcal{O}(\epsilon) = \frac{1}{z \alpha(z)} + \mathcal{O}(\epsilon) \,,
\end{align}
where $\alpha(z)$ is defined in \cref{eq:def yukawa coupling}. 
We therefore choose the new integral $J_2=z \alpha(z) \partial_z J_1$. 
Proceeding in the same way, we take the derivative of $J_2$ and normalize it to unity at $\epsilon=0$ on $C_2$ to define $J_3$, and differentiate $J_3$ and normalize it to unity at $\epsilon=0$ on $C_3$ to define $J_4$. 
This yields the basis
\begin{equation}\label{eq:canbasbanana}
    \begin{aligned}
        J_1 &= I_\text{Ban4L}/\pi_0         \, ,&
        J_2 &= z \alpha(z) \partial_z J_1   \, , \\
        J_3 &= -\frac{(z-1) (9 z-1) (25 z-1) \pi_0(z)^2}{z \alpha(z)^2} \partial_z J_2          \, ,&
        J_4 &= z \alpha(z)  \partial_z J_3  \, .
    \end{aligned}
\end{equation}
Let us notice here, that the leading singularity of $\partial_z J_1$ along $C_1$ and $\partial_z J_3$ along $C_3$ are equal. This is related to the special form of a Calabi-Yau operator (we refer to \cite{Duhr:2022dxb} for more details, in particular eqs. (5.6) and (5.8) in there). The basis \cref{eq:canbasbanana} has normalized leading singularities at $\epsilon=0$ in the sense of \cite{Forner:2026vby}. 
As pointed out there, one finds non-vanishing leading singularities $\text{LS}[J_i]_{C_j}$ for $i\leq j$, e.g., $\text{LS}[J_1]_{C_1}=\pi_1/\pi_0$.
Evaluating the whole basis $\vec J$ on the cycles $C_i$ gives
\begin{equation}
	\text{LS}[\vec J]_{C_i}	=	\begin{pmatrix} 
    1   &   \frac{\pi_1}{\pi_0} &   \frac{\pi_2}{\pi_0} &   \frac{\pi_3}{\pi_0} \\
    0   &   1                   &   \frac{\pi_0\pi_2'-\pi_0'\pi_2}{\pi_0\pi_1'-\pi_0'\pi_1}                 &   \frac{\pi_1\pi_2'-\pi_1'\pi_2}{\pi_0\pi_1'-\pi_0'\pi_1}                 \\
    0   &   0                   &   1                   &   \frac{\pi_1}{\pi_0} \\
    0   &   0                   &   0                   &   1
    \end{pmatrix}	+ \mathcal O(\epsilon) \, ,
\label{eq:Wuni}
\end{equation}
which is a unipotent matrix containing the non-constant leading singularities. Generalizations of this construction were discussed in~\cite{Duhr:2022dxb, Duhr:2025lbz}. As was argued in \cite{Forner:2026vby}, the unipotent leading singularities in \cref{eq:Wuni} are correctly normalized in a generalized notion of the polylogarithmic case.

Note also that, as pointed out in~\cite{Forner:2026vby}, the rotation matrix from the initial derivative basis to this new basis is exactly equal to the inverse semi-simple part of the Wronskian matrix, and thus equivalent to the first step in the method proposed in~\cite{Gorges:2023zgv,Duhr:2025lbz}.
After this overall normalization, we account for the weight drops introduced by the various
derivatives by the $\epsilon$-rescaling $T_\epsilon=\text{diag}(1,\epsilon^{-1},\epsilon^{-2},\epsilon^{-3})$. Then, we see that only entries below the diagonal are not yet $\epsilon$-factorized. This can be interpreted as non-constant leading singularities at higher orders in $\epsilon$. The fact that these extra leading singularities can be removed by a finite number of $\epsilon$-independent rotations is a consequence of the choice of our starting basis, which is built out of derivatives from our starting integral $J_1$ that is fully canonical.

While it is in principle possible to construct this rotation by integrand analysis only, the simplest way to obtain it in practice is by working directly at the level of the differential equations by the use of an ansatz. This requires introducing the four new functions $G_1,\dots,G_4$ defined in~\cref{app:Gfunctions}. Note that they are built from $\pi_0$, $\alpha$ and rational functions only.

\subsection{The eyeball graph}
\label{sec:eyeball}

\begin{figure}[t]
    \centering
    \includegraphics[width=0.3\linewidth]{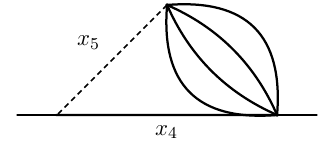}
    \caption{A four-loop eyeball graph, corresponding to sector 1265. It is polylogarithmic on the maximal cut, but $\epsilon$-factorizing its coupling to the banana sector requires a new transcendental function.}
    \label{fig:eyeball}
\end{figure}

The banana sector is the only one where the Calabi-Yau threefold is visible on the maximal cut. 
However, the same geometry also enters sectors whose maximal cuts are much simpler. 
As the simplest example, let us now turn to the eyeball graph shown in \cref{fig:eyeball}, which corresponds to sector 1265 and has a single master integral. 
We consider the corner integral $I_{\text{eye}}=I_{1 0 0 0 1 1 1 1 0 0 1 0 0 0}$ in two dimensions and introduce the Baikov parametrization
\begin{align}
    I_{\text{eye}} \propto \int \dd x_4 \dd x_5 \frac{z}{x_4 x_5 \sqrt{z^2 \left(x_4-x_5+1\right){}^2-2 z \left(x_4+x_5+1\right)+1}}\,  I_{\text{Ban3L}}(x_5) \, .
\end{align}
Here, we used the fact that this integral can be interpreted as a bubble with one massive propagator $x_4$ and a massless propagator $x_5$, dressed with a three-loop equal-mass banana integral, which we denoted by $I_{\text{Ban3L}}(x_5)$.
We recall here that $I_{\text{Ban3L}}$ itself requires three integrations $x_1, x_2, x_3$ in the loop-by-loop approach. 
We ignore the part of the three-loop banana graph integrand which can be expressed as a series of logarithmic integrations, and focus on its residue 
which reads
\begin{align} \label{eq:integrand eyeball}
    I_{\text{eye}} \propto\int \dd x_2 \dd x_3 \dd x_5 \frac{1}{x_5}\frac{1}{\sqrt{x_2 (x_2-4) Q(x_2,x_3) Q(x_3, x_5) Q(x_5, 1/z)}}\,,
\end{align}
which exposes the square root of the polynomial of \cref{eq:polyCY3}, 
and an extra pole in $x_5$. Note that we did not need to perform any cuts to arrive here, so this result is exact.

To analyze the leading singularities, we consider the independent closed contours in $x_5$. If we pick the contour around the pole, which corresponds to the maximal cut of the integral, the square root degenerates and we are left with
\begin{align}
\label{eq:integrand eyeball MC}
    \int \dd x_2 \dd x_3 \frac{z}{(1-z)  \left(x_3-1\right) \sqrt{x_2} \sqrt{x_2-4}\sqrt{-x_2^2+2 \left(x_3+1\right) x_2-\left(x_3-1\right){}^2}}\,.
\end{align}
From here, we can take the residue at $x_3=1$, which leaves two poles at $x_2=0$ and $x_2=4$. Due to the global residue theorem, these are not independent and we need to consider only one of them. Choosing $x_2=0$, we recover the leading singularity of the maximal cut,
\begin{align}
    \text{LS}[I_{\text{eye}}]_{x_5=0,\, x_3=1,\, x_2=0} = \frac{z}{1-z}\,.
\end{align}
This tells us that we need to rescale the integral by $(1-z)/z$ to make it canonical on the maximal cut and, 
from now on, we will consider this rescaled integral.
We are not done yet, as there are other contours we need to consider to make sure all leading singularities are constant.

Instead of taking the residue, 
we can choose as an integration contour one of the cycles of the Calabi-Yau geometry, $C_0,\dots,C_3$. For our analysis, it is enough to consider the holomorphic cycle $C_0$ explicitly. 
The leading singularity on $C_0$ is a generalized third-kind integral on the Calabi-Yau geometry and needs to be subtracted, which is the origin of a new function necessary for the $\epsilon$-factorization. As proposed in~\cite{Forner:2026vby} the new function can be predicted in four steps:
\begin{enumerate}
  \item compute the local series expansion of the integrand on the holomorphic cycle of the subsector geometry;
  \item find a differential equation for this series, using an ansatz guided by the geometries involved;
  \item simplify it by shifting the function by multiples of the periods;
  \item solve it in terms of iterated integrals.
\end{enumerate}
For the eyeball graph, the series expansion of $\frac{1-z}{z}\text{LS}[I_{\text{eye}}]_{C_0}$ can be obtained from \cref{eq:integrand eyeball} in the following way. First, we expand the integrand around $z=0$, which simplifies the factor $Q(x_5,1/z)$ in the square root. The holomorphic period then corresponds to the contour that encircles the poles at $x_5=\infty$, $x_3=\infty$ and $x_2=\infty$, i.e., it can be computed simply by taking consecutive residues. Let us denote this object by $G(z)=-\frac{1-z}{z}\text{LS}[I_{\text{eye}}]_{C_0}$ (such that the function $G(z)$ will be the object we need to add to remove the leading singularity). We find
\begin{align}
    G(z) = -z -7 z^2 -65 z^3 -791 z^4 -  11477 z^5 + \mathcal{O}(z^6) \,,
\end{align}
displaying only the first five orders here. Going to sufficiently high orders, one can use this series expansion to find a differential equation for $G(z)$. Since we know that $G(z)$ comes from a sector which is polylogarithmic on the maximal cut and couples to the Calabi-Yau geometry of the banana sector, it is reasonable to make an ansatz of the form of a first-order differential equation for $G$ with an inhomogeneity consisting of the period $\pi_0$ and its derivatives. Indeed, we find
\begin{align}
    5 (1-z) z G'(z)+z (7+5 z) \pi_0'(z)-2 \pi_0(z)= 0\,.
\end{align}
While it is not strictly necessary, we can redefine $G(z)$ by a shift to remove either $\pi_0$ or $\pi_0'$ from this equation. Choosing the latter, we define 
\begin{equation}
G(z) = G_5(z) -\frac{\left(7+5z\right) }{5 \left(1-z\right)}\pi_0(z) 
\end{equation}
and obtain
\begin{align}
     G_5 '(z)  = \frac25\frac{1+5 z }{ (1-z)^2 z}\pi_0(z)\,.
\end{align}
This tells us that, to remove this extra leading singularity given by $G$, we should construct the integral
\begin{align}
    \frac{1-z}{z} I_{\text{eye}} + G(z) J_1 = \frac{1-z}{z} I_{\text{eye}}+ \left(G_5(z) - \frac{7+5z }{5 \left(1-z\right)}\pi_0(z) \right) J_1\,,
\label{eq:eyecan}
\end{align}
where $J_1$ is the first canonical integral of the banana sector defined in \cref{eq:canbasbanana}, and
\begin{align}
    G_5(z) = \int^z \dd u \, \frac25  \frac{1+5 u }{ (1-u)^2 u}\pi_0(u)
\end{align}
is given as an (iterated) integral over $\pi_0$ and a rational function.
We stress here that the lower integration boundary is not important since it only shifts the integral by a constant multiple of the canonical integral $J_1$ and can be reabsorbed in the overall boundary constants needed to 
fix the solution of the final differential equations. As a practical consequence, the boundary constant can be fixed arbitrarily also for the numerical evaluation of this function~\cite{Czakon:2013goa}.
One can verify that the integral in~\cref{eq:eyecan} indeed satisfies a canonical differential equation.

\subsection{The double eyeball graph}\label{sec:eyeball2}

\begin{figure}[t]
    \centering
    \includegraphics[width=0.3\linewidth]{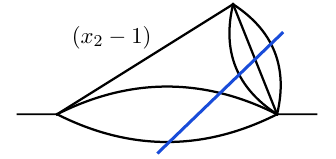}
    \caption{The four-loop double-eyeball graph, which corresponds to sector 1520. Here, we explicitly discuss the cut indicated by the blue line, which isolates the top sector and its coupling to the equal-mass banana.}
    \label{fig:doubleeyeball}
\end{figure}

In the eyeball graph discussed in the previous subsection, the top-sector itself is polylogarithmic on the maximal cut. As an example where two non-trivial geometries couple, we now consider sector $1520$, the equal-mass ``double eyeball'' graph, shown in \cref{fig:doubleeyeball}, which is elliptic on the maximal cut. This example is similar to the three-loop double eyeball graph studied in~\cite{Forner:2026vby}, and we follow the same steps as in the previous subsection.

This sector has two master integrals, but we will here only discuss one of them. Taking the cut indicated by the blue line in  \cref{fig:doubleeyeball} allows us to focus on the most interesting part, which is the coupling of this sector to the four-loop equal-mass banana sector. We denote this cut integral by $I_{\text{2eye}}= \text{Cut}_{\text{CY3}}\, I_{00001111101000}$.
First, we derive the Baikov representation
\begin{align}\label{eq:I2eye}
    I_{\text{2eye}} \propto  \int\dd x_1 \dd x_2 \dd x_3 \frac{1}{(x_2-1)\sqrt{x_1 (x_1-4) Q(x_1,x_2) Q(x_2, x_3) Q(x_3, 1/z)}}\,.
\end{align}
Again, this integrand is equal to that of the equal-mass banana, up to an extra pole that corresponds directly to the additional (uncut) propagator in the graph. Taking the residue at $x_2=1$, which projects  onto the maximal cut, we find the holomorphic form of the elliptic sunrise geometry. 
Indeed, this sector is elliptic on the maximal cut, and this analysis tells us that we should rescale the integral as $I_{\text{2eye}}/\varpi_0(z)$ to normalize its leading singularity on the maximal cut. 

To investigate its coupling to the banana graph, we derive a series expansion for $\text{LS}[I_{\text{2eye}}]_{C_0}$ around $z=0$, analogous to the previous subsection. 
As for that case, the expansion around $z=0$ is easy to compute since it collapses the holomorphic cycle to iterative residues. 
We denote $\bar G(z)=-\text{LS}[I_{\text{2eye}}]_{C_0}$ (such that the function $\bar G(z)/\varpi_0(z)$ will be the object we need to add to remove the leading singularity) and find\begin{equation}
    \bar G(z) = -z^2 -12 z^3 -145 z^4 - 1904 z^5 + \mathcal{O}(z^6)\,.
\end{equation}
Since $\bar G(z)$ corresponds to a sector with the elliptic geometry of the sunrise on the maximal cut and a coupling to the Calabi-Yau threefold of the banana, it is reasonable to try as an ansatz a second-order inhomogeneous differential equation of the form
\begin{align}
    (1-z)(1-9 z)z^2 \mathcal{L}_2 \bar G(z) + b_1(z)\pi_0(z) + b_2(z)\pi_0'(z) + b_3(z)\pi_0''(z) + b_4(z)\pi_0'''(z) = 0\,.
\end{align}
Here, $\mathcal{L}_2$ is the second-order differential operator of the sunrise geometry defined in \cref{eq:deqsunperiods}, $b_i(z)$ are rational functions, and we included a prefactor in front of $\mathcal{L}_2$ for convenience. 
Indeed, on the level of the series expansion of $\bar G(z)$ we find the differential equation
\begin{align}
    (1-z)(1-9 z)z^2 \mathcal{L}_2 \bar G(z) +\frac{3z}{10} \pi_0(z) +\frac1{10}(7-45z) z^2 \pi_0'(z) +\frac{3}{10} (7-15 z) z^3 \pi_0''(z) = 0\,.
\end{align}
Next, we divide $\bar G(z)$ by the leading singularity of the maximal cut, $\varpi_0(z)$, and then use the freedom to shift it to remove $\pi_0''(z)$ from the differential equation. 
In total, we define 
\begin{equation}
\bar G(z) = \varpi_0(z) G_7(z)-\frac3{10}\frac{(7-15 z)z }{(1-z)(1-9z)}\pi_0(z)\,,
\end{equation}
such that $G_7(z)$ satisfies
\begin{equation}
    \begin{split}
        &10 (1-z)^3 (1-9 z)^3z \varpi_0(z) G_7''(z)\\
    &\quad+ \left[20 (1-z)^3 (1-9 z)^3z \varpi_0'(z)-10 (1-z)^2 (1-3 z) (1+3 z) (1-9 z)^2 \varpi_0(z)\right]G_7'(z)\\
    &\quad-2  \left(7-10 z+99 z^2\right)(1-z)  (1-9 z)z \pi_0'
    +3 \left(1-54z+52z^2+54z^3+459z^4\right) \pi_0 = 0\,.
    \end{split}
\end{equation}
We highlight that this second-order differential equation is particularly simple since it involves only the first and second derivative of $G_7(z)$ and is therefore equivalent to a first-order equation 
for the function $f(z) = G_7'(z)$. As it's easy to see, it homogeneous part reads
\begin{align}
    \frac{f_{\text{hom}}'(z)}{f_{\text{hom}}(z)} = -2\frac{ \varpi_0'(z)}{\varpi_0(z)}+\frac{(1-3 z) (1+3 z)}{(1-z) (1-9 z)z}
\end{align}
and can then easily be integrated giving
\begin{align}
    f_{\text{hom}}(z) = -c\frac{z}{(1-9 z) (1-z) \varpi_0(z)^2}
\end{align}
with $c$ an integration constant. 
From this, we can solve the full inhomogeneous differential equation using standard variation of constants. 
The solution is
\begin{equation}
f(z) = -G_6(z)\frac{z}{(1-9 z) (1-z) \varpi_0(z)^2}\,,
\end{equation}
where
{\small
\begin{align}
    G_6(z)&=\int^z \dd u  \left( 
    \frac3{10} \frac{ \left(1-54u+52u^2+54u^3+459u^4\right)  }{ (1-u)^2 (1-9 u)^2 u^2}\pi_0(u)
    -\frac{7-10 u+99 u^2 }{5 (1-u) (1-9 u)u}\pi_0'(u)\right)\varpi_0(u)\,.
\end{align}}
Finally, we have
\begin{align}
    G_7(z)=\int^z \dd u\, f(u)= -\int^z \dd u \, \frac{u}{(1-u) (1-9 u) \varpi_0(u)^2}G_6(u) \,,
\end{align}
and we construct the integral
\begin{align}
\label{eq:canonical int double eyeball}
   \frac{I_{\text{2eye}} + \bar G(z)J_1}{\varpi_0(z)}=  \frac{I_{\text{2eye}}}{\varpi_0(z)} + \left(G_7(z) -\frac3{10}\frac{ (7-15 z)z }{(1-z)(1-9z)  \varpi_0(z)}\pi_0(z)\right)J_1 \,,
\end{align}
where $J_1$ is the first canonical integral of the banana sector defined in \cref{eq:canbasbanana}. One can verify on the level of the differential equations that the integral defined by~\cref{eq:canonical int double eyeball} is canonical. We stress that the derivation of~\cref{eq:canonical int double eyeball} was entirely based on integrand analysis.

Compared to the eyeball graph of~\cref{sec:eyeball}, 
the new $G$-function required in this case is  a two-fold iterated integral, 
which mixes the periods of the two geometries. T
he inner integration kernel of $G_6(z)$ contains products of $\varpi_0(z)$ with $\pi_0(z)$ and $\pi_0'(z)$, 
while the outer one in $G_7(z)$ comes from the normalization of the elliptic maximal cut.
By obvious generalization of the methods described here, this 
analysis can be extended beyond this cut. 
One can also perform a similar analysis for the second master integral in the sector, 
which corresponds to the second-kind form on the maximal cut. 
Since a good starting point for the second integral is the derivative of \cref{eq:canonical int double eyeball}, $G_6(z)$ appears very naturally in this construction. 
However, it also leads to the functions $G_8(z)$ and $G_9(z)$ in addition, which are defined in~\cref{app:Gfunctions}. 
This analysis is more involved and we do not include it here for brevity, while we stress that is it often simpler to 
derive the corresponding functions directly from the system of differential equations.

\subsection{The cat-eye graph}
\label{sec:cateye}

As a last explicit example, let us discuss the cat-eye graph of \cref{fig:cateye}, whose sector consists of five master integrals. In contrast to the previous examples, one of the most interesting couplings now occurs already on the maximal cut, namely between two different elliptic curves related to two different parametrizations of the sunrise family. We therefore consider the maximal cut, which allows us to ignore the couplings to the Calabi-Yau threefold
and focus on the elliptic geometry of the sector itself.

\begin{figure}[t]
    \centering
    \begin{subfigure}[b]{0.5\textwidth}
         \centering
         \includegraphics[width=0.6\textwidth]{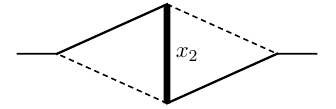}
        \caption{}
    \label{fig:cateye a}
     \end{subfigure}%
     \begin{subfigure}[b]{0.27\textwidth}
         \centering
         \includegraphics[width=0.5\textwidth]{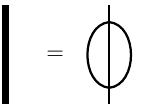}
         \caption{}
         \label{fig:cateye b}
     \end{subfigure}
    \caption{To parametrize the cat-eye graph in \cref{fig:cateye}, we first consider a two-loop kite graph (a) with one generalized propagator represented by a thick line, which we then choose as the two-loop sunrise graph (b).}
        \label{fig:cateyeBaikov}
\end{figure}

We construct a particularly simple and convenient loop-by-loop Baikov parametrization by decomposing the cat-eye graph in \cref{fig:cateye} into two building blocks, a kite and a sunrise topology.
To do so, we first consider the auxiliary two-loop kite family in \cref{fig:cateye a}, with Baikov variables
\begin{equation}
    \begin{aligned}
    x_2 &= k_1^2\,,& x_3 &= k_2^2 - m^2\,, &x_4 &= (p-k_2)^2\,,\\
        x_5 &= (k_1 - k_2)^2\,, &x_6 &= (p+k_1-k_2)^2 - m^2\,. &&
\end{aligned}
\end{equation}
Here, however, we allow the thick line in the middle to be a generalized propagator that may, in general, carry an arbitrary function $f(x_2)$ of the Baikov variable $x_2$, which parametrizes the squared momentum flowing through the line. 
The standard kite integral would simply correspond to the choice $f(x_2)=1/x_2$. 
Instead, to parametrize the cat-eye graph, the dressed propagator must be chosen as the two-loop equal-mass sunrise graph, as illustrated in \cref{fig:cateye b}. 
We denote it by $f(x_2)=\text{Sun}_{n_1}(x_2)$, where we also allow for a generic numerator insertion in the Baikov variable $x_1$ of the sunrise. 
Then, the cat-eye integral in $D=2$ dimensions with denominator powers equal to $1$ takes the form
\begin{align}
    I^{\text{cat}}_{n_1,n_2,1,1,1,1}  \propto \int \dd x_2 \cdots \dd x_6 \,\frac{x_2^{n_2}}{x_3 x_4 x_5 x_6\,\mathcal{B}(x_2,\dots,x_6)}\text{Sun}_{n_1}(x_2) 
\end{align}
with another generic numerator insertion $x_2^{n_2}$, and $\mathcal{B}$ is the Baikov polynomial, whose full explicit form is irrelevant for the following discussion.
Then, the maximal cut of the cat-eye integral in $D=2$ dimensions, which we denote by $I^{\text{cat}}_{n_1,n_2}$, is obtained by taking the residues at $x_3=x_4=x_5=x_6=0$ and replacing the sunrise subgraph by its maximal cut. 
We find
\begin{align}
    I^{\text{cat}}_{n_1,n_2} \propto \int \dd x_1 \dd x_2 \,\frac{z\,x_1^{n_1} x_2^{n_2}}{\left(x_2-(2-1/z)\right) \left( x_2-z\right)\sqrt{P_4(x_1;x_2)}}\,,
\end{align} 
where we have again set $m^2=1$, and $P_4$ is the fourth-order polynomial of \cref{eq:sunrisecurve}, defining the elliptic geometry of the sunrise graph.
In addition to the branch cuts from the square root, the integrand showcases two distinct poles in $x_2$.
Each pole gives rise to an independent residue, which we can isolate by considering the combinations
\begin{equation}\label{eq:cateye first kind ints}
    \begin{split}
        &\frac{1}{z} I^{\text{cat}}_{0,1} - I^{\text{cat}}_{0,0} \propto \int \dd x_1 \dd x_2 \, \frac{1}{\left(x_2-(2-1/z)\right) \sqrt{P_4(x_1;x_2)}}\,, \\
    & \frac{1}{z}\left( I^{\text{cat}}_{0,1}-(2-1/z)I^{\text{cat}}_{0,0}  \right) \propto \int \dd x_1 \dd x_2 \, \frac{1}{\left( x_2-z\right) \sqrt{P_4(x_1;x_2)}}\,.
    \end{split}
\end{equation}
Taking the corresponding residues in $x_2$, we find
\begin{align}
    & \frac{1}{z} I^{\text{cat}}_{0,1} - I^{\text{cat}}_{0,0} \propto \int \dd x_1 \, \frac{1}{\sqrt{P_4(x_1;2-1/z)}} \, ,
    \nonumber \\
    & \frac{1}{z}\left( I^{\text{cat}}_{0,1}-(2-1/z)I^{\text{cat}}_{0,0}  \right) \propto  \int \dd x_1 \, \frac{1}{\sqrt{P_4(x_1;z)}} \,,
\end{align}
which uncovers the families of elliptic curves defined in \cref{eq:defellcurvecateye}.
Due to the presence of these two elliptic curves, the maximal cut at $\epsilon=0$ naturally decomposes into blocks of sizes $2+2+1$. 
The $\epsilon$-factorization of the two blocks of size $2$, which correspond to shifted versions of the sunrise graph, works completely analogously to the latter, using the periods $\varpi_0^{\text{(shift)}}$ and $\varpi_0^{\text{(inv)}}$ defined in \cref{sec:elliptic}, respectively. We then start by choosing the two integrals in \cref{eq:cateye first kind ints} and their derivatives as a starting basis, mapping them from the auxiliary integral family to the one defined in \cref{eq:intfam}. However, there remains a non-$\epsilon$-factorized 
coupling between these two blocks. To factorize it, we need to introduce a new transcendental function
\begin{align}
    G_{10}(z) = \int^z \dd u \, \frac{ 2 (5+7 u)u {\partial_u\varpi_0^{\text{(shift)}}}(u)-\varpi_0^{\text{(shift)}}(u)}{u}\varpi_0^{\text{(inv)}}(u)\,,
\end{align}
whose integration kernel involves the periods of both elliptic curves at the same time.
We stress once more that this function can in principle be determined by integrand analysis or, equivalently,
by imposing that the system of differential equations is $\epsilon$-factorized,

For the fifth integral in the sector, we choose a Gram determinant $G(k_1,k_2,p)$ in the numerator, as proposed in~\cite{Duhr:2022dxb} for the closely related ice cone family. Behind this choice is the fact that this integral has at most single poles and is evanescent in $d=2$ space-time dimensions. The fact that it has at most single poles can be easily seen by studying its integrand close to $d=2$ dimensions employing the same representation as in~\cref{fig:cateyeBaikov}. There, the integrand of the sunrise graph has at most single poles close to $d=2$, while the Gram determinant $G(k_1,k_2,p)$ shifts the remaining effective 2-loop kite to $d=4$ dimensions, where the kite itself has at most single poles.
This immediately yields a canonical integral and completes the basis on the maximal cut. The list of our precanonical integrals for this sector and all remaining ones can be found in \cref{app:precan}.

\vspace{0.5em}
Proceeding along the lines of the examples above for all sectors, we obtain a canonical basis for all $160$ master integrals, which satisfies an $\epsilon$-factorized differential equation of the form of~\cref{eq:candeq}. The entries of $B(z)$ are built from rational functions, the square roots $r_1,\dots,r_4$, the periods $\pi_0$, $\alpha$, $\varpi_0$, $\varpi_0^{\text{(shift)}}$, $\varpi_0^{\text{(inv)}}$ and their derivatives, and the functions $G_1,\dots,G_{27}$. 
We checked explicitly that, expanding the differential equation matrix around any of its singularities, there are only simple poles and no logarithms. 
In the next section, we use this differential equation and its series expansions to obtain our results for this family.

\section{Results}
\label{sec:results}

With the canonical basis of \cref{sec:can basis} at hand, we can now discuss our results for the $160$ master integrals. Our main findings can be summarized as follows:
\begin{itemize}
  \item The canonical differential equation can be written in terms of $100$ independent integration kernels. Only $13$ of them are $\mathrm{d}\!\log$-forms, while the remaining $87$ involve the periods of the elliptic curve, of the Calabi-Yau threefold, or both (\cref{sec:analyticresults}).
  
  \item At the MUM point $z=0$, i.e., for $p^2\to\infty$, all boundary constants up to weight seven can be expressed in terms of multiple polylogarithms with letters $0$ and $1$ evaluated at $1/2$, which are also known as alternating multiple zeta values. As expected, no trace of the elliptic curve or of the Calabi-Yau threefold survives in this limit (\cref{sec:mum}).
  
  \item At the on-shell point $z=1$, i.e., for $p^2 = m^2$, we recover the classes of transcendental numbers known from the four-loop $(g-2)$ of the electron~\cite{Laporta:2017okg}. These are alternating multiple zeta values, polylogarithms evaluated at sixth roots of unity, and non-polylogarithmic constants. In addition, we find two new independent constants at weight seven (\cref{sec:onshell}). We stress that the constants encountered here already cover all planar non-polylogarithmic constants appearing in the calculation of the four-loop QED corrections to the $(g-2)$~\cite{Laporta:2017okg}.
  
  \item Using the canonical differential equation, all master integrals can be evaluated numerically everywhere in the kinematic space in fractions of a second (\cref{sec:numerical}).
\end{itemize}

\subsection{General analytic solution in terms of iterated integrals}
\label{sec:analyticresults}

Expanding the canonical basis in the dimensional regulator
\begin{equation}
\vec J(z,\epsilon) = \sum_{k\geq 0} \epsilon^k \vec J^{(k)}(z)\,,
\end{equation}
we see that~\cref{eq:candeq} implies
\begin{equation}\label{eq:itsol}
  \vec J^{(k)}(z) = \vec J^{(k)}(z_0) + \int_{z_0}^{z} \dd x\, B(x)\, \vec J^{(k-1)}(x)\,,
\end{equation}
where we normalized all master integrals such that $\vec J^{(-1)} = 0$. 
Hence, at each order in $\epsilon$, the master integrals are given by iterated integrals over the entries of $B(z)\mathrm{d}z$, and they are fully determined once the boundary values $\vec J^{(k)}(z_0)$ are known. If $z_0$ is a singular point of the differential equation, \cref{eq:itsol} is to be understood with the usual tangential base point regularization, which we can assume since all our differential forms have at most simple poles.
The functions appearing in our result are therefore determined by the integration kernels contained in $B(z)\mathrm{d}z$, whereas the transcendental numbers follow from the boundary constants.

To make this function space explicit, we write the entries of $B(z)\mathrm{d}z$ as $\mathbb{Q}$-linear combinations of kernels which are linearly independent up to total derivatives. 
We verify this employing high-order series expansions for the kernels.\footnote{We thank Benjamin Sauer for sharing a private code to efficiently check for linear relations among series expansions.}
Importantly, if one had initially introduced a redundant set of $G$-functions, this step would 
also uncover relations among them. We have checked that this is not the case for our set of $G$-functions. 
As already announced above, through this procedure we find that our differential equation matrix can be written in terms of $100$ kernels, out of which $13$ are $\mathrm{d}\!\log$-forms.

To describe our kernels, we start by defining the four square roots
\begin{align}
    r_1(z) = \sqrt{1-z}\,, \quad r_2(z) = \sqrt{1-9z}\,, \quad r_3(z) = \sqrt{1+3z}\,, \quad r_4(z) = \sqrt{1+z}\,.
\end{align}
The $\mathrm{d}\!\log$-forms can be split into a set of $8$ rational and $5$ algebraic kernels, which are
\begin{align}
    \left\{\frac{25}{25 z-1},\frac{9}{9 z-1},\frac{1}{z-1},\frac{1}{z-9},\frac{1}{z},\frac{1}{z+1},\frac{3}{3 z+1},\frac{7}{7 z+1}\right\} \, ,
\end{align}
and
\begin{align}
    \left\{\frac{1}{r_1 r_2},\frac{1}{z\, r_1 r_2},\frac{1}{r_1 r_3},\frac{1}{z\, r_1 r_3},\frac{1}{z\, r_1 r_4}\right\}\,.
\end{align}
The latter all integrate to logarithms, since
\begin{align}
    \int \dd z \,\frac{1}{\sqrt{a-b z} \sqrt{c-d z}} = \frac{1}{\sqrt{b}\sqrt{d}}\log\left(\frac{1-\frac{\sqrt{b} \sqrt{c-d z}}{\sqrt{d} \sqrt{a-b z}}}{1+\frac{\sqrt{b} \sqrt{c-d z}}{\sqrt{d} \sqrt{a-b z}}}\right)\,.
\end{align}
Notice that the kernels with an additional factor $1/z$ are mapped to the same form by the change of variables $y=1/z$.

The remaining $87$ kernels depend on transcendental functions of elliptic, Calabi-Yau or mixed type. They are expressed in terms of the periods $\varpi_0$, $\varpi_0^{\text{(shift)}}$, $\varpi_0^{\text{(inv)}}$ of the elliptic curves, the functions $\pi_0$ and $\alpha$ of the Calabi-Yau threefold introduced in \cref{sec:general}, their derivatives, and the $27$ functions $G_1,\dots,G_{27}$ of \cref{app:Gfunctions}.  We stress here that, by 
series expanding the differential equations around any singular point, one can easily confirm 
that the connection matrix has only simple poles and, thanks to the choices made in \cref{sec:general}, that its expansion is free of logarithms.

In total, the singular points of the differential equation are
\begin{equation}
    z\in\{0, \tfrac{1}{25}, \tfrac{1}{9}, \tfrac12, 1, 9, -\tfrac17, -\tfrac13, -1, \infty\},
\end{equation}
where the point $z=1/2$ appears only through the non-polylogarithmic kernels, and one can verify that at each singular point at most logarithmic singularities appear.

\subsection{Series solutions and boundary constants}
\label{sec:series and boundaries}

To fix the boundary values in \cref{eq:itsol} and to study the transcendental numbers appearing in our results, we compute series expansions of the canonical master integrals around two singular points.
These are the MUM point $z=0$, where the structure is the simplest, and the physically most interesting on-shell point $z=1$, related to the electron $(g-2)$.

At both points, we proceed in a similar way. 
Starting from the local expansions of the periods given in \cref{sec:general}, 
we derive series expansions for all $G$-functions, and thus for all kernels in the canonical differential equation matrix. This allows us to derive a high-order series-expansion representation for the general solution of the canonical differential equation around the expansion point, which still depends on a set of integration constants. 
To fix them, we evaluate the master integrals of our starting basis numerically with high precision, using \texttt{AMFlow}~\cite{Liu:2017jxz, Liu:2022chg}, at a point close to the expansion point.
Then we rotate these values to the canonical basis, and equate them with our general series solution evaluated at the same point. Solving the resulting system of equations yields high-precision numerical values for the boundary constants, which we finally match to analytic constants with the PSLQ algorithm, as implemented in \texttt{Mathematica}.

\subsubsection{The large-momentum limit $z=0$}
\label{sec:mum}

Around $z=0$, i.e., for $p^2\to\infty$, we evaluate the starting basis with \texttt{AMFlow} at $z=10^{-3}$ with $300$ digits precision. Since we match at a point with a slight offset from the expansion point, it is essential to derive series expansions to high orders to keep enough precision. 
Here we used series expansions with $200$ orders.

The PSLQ matching additionally requires an educated guess of the transcendental constants expected at the boundary point. At the first few orders in $\epsilon$, i.e., the first few weights, where the matching is still simple, we observe only zeta values and multiple polylogarithms with letters $0$ and $1$ evaluated at $1/2$. We conjecture that this remains true at higher weights. Then, to construct a basis of constants at a given weight, we generate lists of all possible products of such constants, evaluate them numerically with \texttt{GiNaC}~\cite{Vollinga:2004sn} through \texttt{PolyLogTools}~\cite{Duhr:2019tlz}, and reduce them to a minimal basis. We always match the real parts first. The constants appearing in the imaginary parts at weight $n$ are then given by $\pi$ times the constants that appeared in the real parts at weight $n-1$.
With these bases, we were able to match all boundary constants up to weight $7$, which is the highest weight we expect to contribute to the electron self-energy. In total, the constants that appear (and products thereof) in the real parts up to weight $7$ are
\begin{align}
    \Big\{&\log(2), \pi^2, \zeta_3,\text{Li}_4(1/2), \zeta_5,   \text{Li}_5(1/2), \text{Li}_6(1/2),G({0, 0, 0, 0, 1, 1}; 1/2),\nonumber\\
    &\qquad \zeta_7, \text{Li}_7(1/2),G({0, 0, 0, 0, 0, 1, 1}; 1/2), G({0, 0, 0, 0, 1, 0, 1}; 1/2)\Big\} \, .
\end{align}
In this list, it is possible to trade the constant $G({0, 0, 0, 0, 1, 1}; 1/2)$ for the combination $\text{Li}_6(1/4)-\frac{16}{243}\text{Li}_6(-1/8)$ and $G({0, 0, 0, 0, 1, 0, 1}; 1/2)$ for $\text{Li}_7(1/4)-\frac{32}{729}\text{Li}_7(-1/8)$. However, we did not manage to replace $G({0, 0, 0, 0, 0, 1, 1}; 1/2)$ by combinations of classical polylogarithms.

It is interesting to note that, although the canonical differential equation involves the periods of the elliptic curves and of the Calabi-Yau threefold, these geometries leave no trace in the boundary constants at $z=0$. This is in line with the fact that $z=0$ is a point of maximal unipotent monodromy for all geometries in our problem, where all periods degenerate to powers of logarithms.

\subsubsection{The on-shell point $z=1$ and relation to the $(g-2)$ of the electron}
\label{sec:onshell}

To expand around $z=1$, i.e., the on-shell limit $p^2 = m^2$, we introduce the local variable
\begin{align}
    u = \frac{1}{z}-1 \, ,
\end{align}
and expand around $u=0$.
All square roots, periods, and $G$-functions were defined around $z=0$ as solutions of rational differential equations. To expand around $u=0$, we transform these defining equations to the new local variable $u$ and choose a new set of solutions, following the criteria discussed in \cref{sec:general}.
These are that the canonical differential equation must have only simple poles at $u=0$, its expansion must be free of logarithms, and the transformed quadratic relations~\cref{eq:griffiths} must be satisfied.

Let us illustrate how these criteria constrain the choice of periods in \cref{eq:expansions1}. Since $\alpha$ is computed only from $\pi_0$ and $\pi_1$, it does not contain logarithms with our choice of periods. However, some choices of periods lead to logarithms in the $G$-functions. As an example, consider the function $G_5(z)$ defined in \cref{sec:eyeball} through 
\begin{equation}
    G_5 '(z)  = \frac{2 (5 z+1) }{5 (z-1)^2 z}\,\pi_0(z)\,.
\end{equation}
Transforming to the local variable $u$, we obtain
\begin{align}
    G_5(u) = -\int^u \dd x \, \frac{2(6+x) }{5 x^2}\,\pi_0(x)
\end{align}
up to a boundary constant that plays no role in this discussion. Computing $G_5(u)$ as a series expansion around $u=0$ by expanding the integrand, and assuming the general structure $\pi_0(x) = a + b x + \mathcal{O}(x^2)$, we find
\begin{align}
    G_5(u) = \frac{2}{5} \left(\frac{6 a}{u} - (a+6 b) \log (u)\right) + \mathcal{O}(u)\,.
\end{align}
By setting $a=b=0$, we remove both the simple pole, which stems from a double pole in the integrand of $G_5$, and the logarithm.
Unfortunately, this introduces double poles in the differential equation matrix. 
However, since we only require the absence of logarithms, we can choose $a\neq0$ and $b=-a/6$ to remove the logarithm, exactly as we did in \cref{eq:expansions1}. It should be clear that we are only allowed to choose $b$ in this way because it is not constrained by the quadratic relations in~\cref{eq:griffiths}. 
In general, we thus allow double poles in the integrands of the $G$-functions, as long as they integrate to simple poles.
But we do not allow for a simple pole in the integrand as they produce a logarithm. 
With these choices, all three criteria above can be satisfied at $z=1$. 
It is an interesting question that we postpone to a future publication,
to determine under which general conditions it is possible to remove all logarithms at non-MUM points.

Once the special functions are fixed, general series solutions and numerical boundary values are computed exactly as around $z=0$. From series expansions around $u=0$ and numerical boundary values computed with \texttt{AMFlow}, we obtain the canonical boundary constants to a precision of $350$ digits. Their analytic structure, however, is significantly richer than at $z=0$.
This is expected, since for on-shell kinematics at $u=0$, the integral families for the electron self-energy coincide with those effectively needed for the electron $(g-2)$.
Indeed, we find that the particular ladder integral family considered in this paper shares many master integrals with the four-loop electron $(g-2)$, and we encounter the same classes of non-polylogarithmic numbers observed for the planar contributions there~\cite{Laporta:2017okg}.

At every weight, i.e., at every order in $\epsilon$, we first reduce the boundary constants numerically to a minimal set of basis constants. We choose this basis such that it decomposes into three groups:
\begin{enumerate}

  \item The first group evaluates to alternating multiple zeta values only, and is easily matched with the PSLQ algorithm.
  
  \item The second group consists of the boundary constants for the integrals in sector $1145$. 
  Already at lower weights, we observe that these involve multiple polylogarithms with letters $0, \pm 1$, evaluated at sixth roots of unity. The larger ansatz requires then much higher precision for the matching.  
  To obtain the required precision, we start from the relevant subsystem of canonical differential equations and use 
  it to transport the analytic boundary values from $z=0$ to $z=1$, solving it with series expansions of up to $6700$ 
  orders on $39$ patches. 
  In this way, we obtain the boundary constants of sector $1145$ with $4000$ digits, 
  which suffices to match them to combinations of polylogarithms of the type described above.
  
  \item The third group contains the canonical boundary constants of the integrals that are given in~\cite{Laporta:2017okg} as a basis of non-polylogarithmic constants. In addition, we find two more independent constants at weight $7$ in the same sectors. A possible explanation is that these constants do not contribute to the physical result of the $(g-2)$, and were hence not given in~\cite{Laporta:2017okg}. Alternatively, there might exist further relations, visible only at higher precision. In either case, we do not attempt to match the constants of these sectors to analytic constants at this point, and work with their numerical values instead.
  
\end{enumerate}

\subsection{Numerical evaluation}
\label{sec:numerical}

\begin{figure}[t]
    \centering
    \begin{subfigure}[b]{0.5\textwidth}
         \centering
         \includegraphics[width=\textwidth]{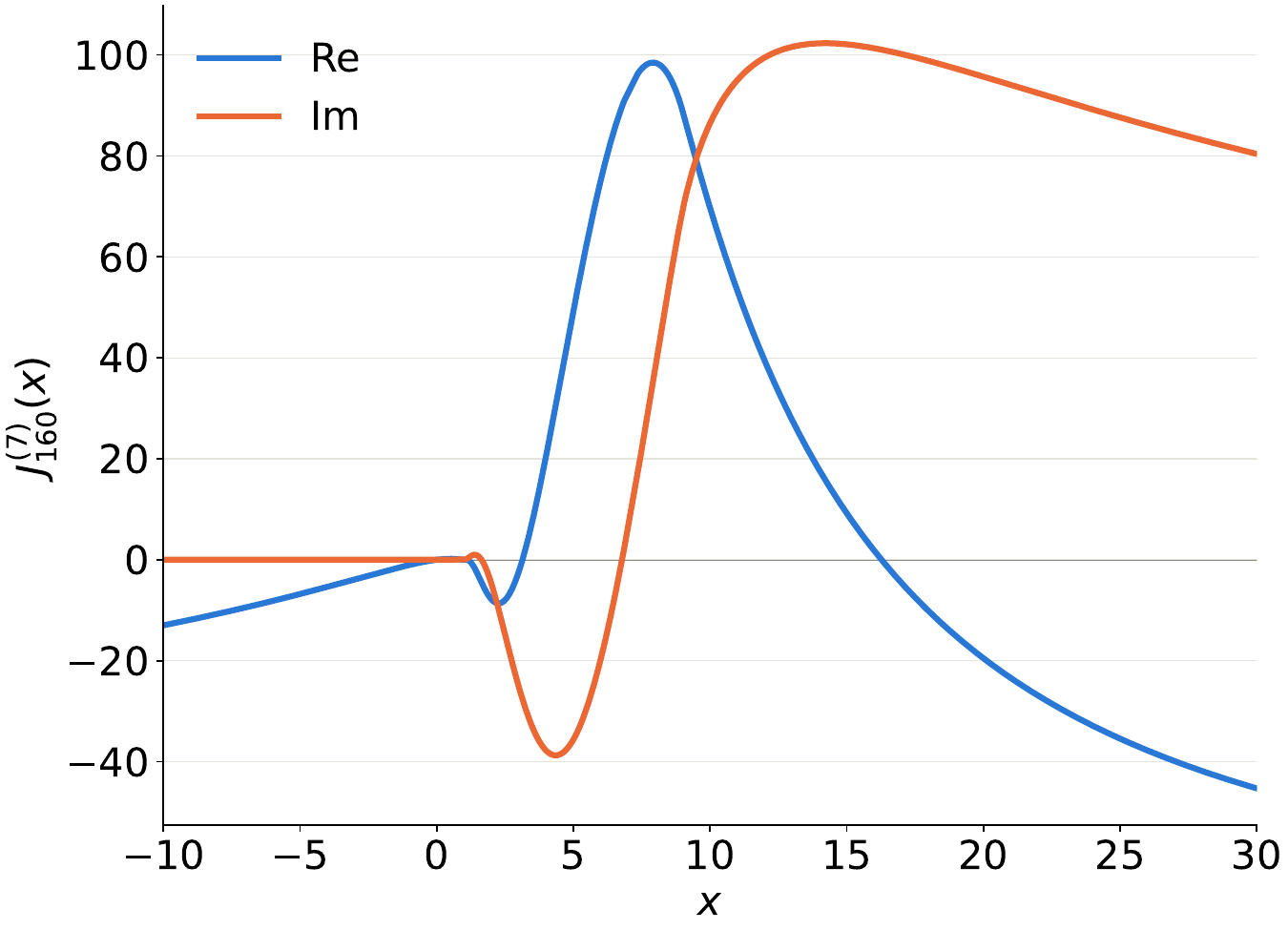}
        \caption{}
    %\label{fig:J160eps7}
     \end{subfigure}%
     \begin{subfigure}[b]{0.5\textwidth}
         \centering
         \includegraphics[width=\textwidth]{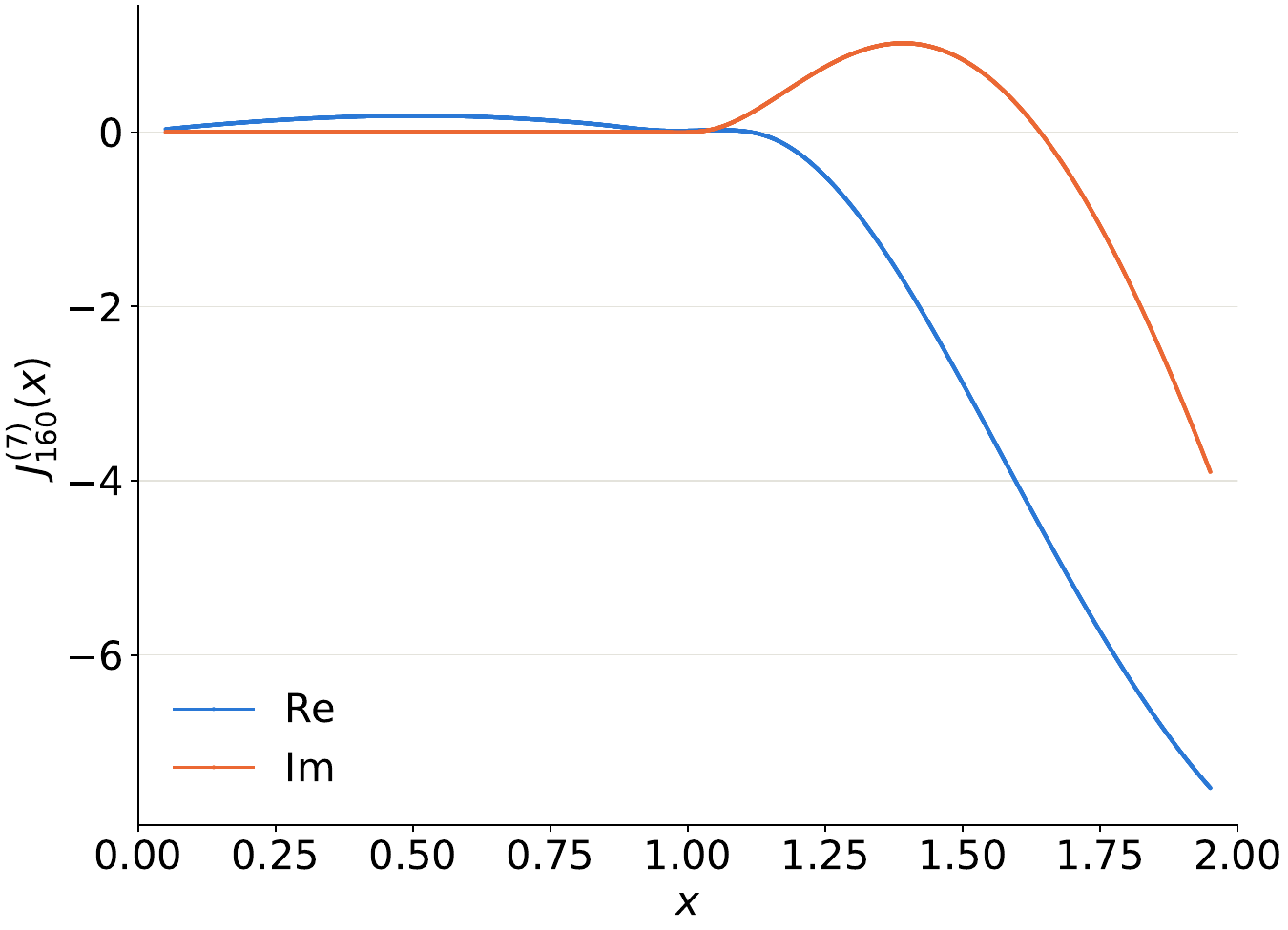}
         \caption{}
         %\label{fig:J160eps7zoom}
     \end{subfigure}
    \caption{Real (blue) and imaginary (orange) parts of the canonical top-sector integral at weight seven, $J_{160}^{(7)}(x)$, in the range $x\in [-10,30]$ (a) and zoomed in on the threshold at $x=1$ (b).}
        \label{fig:J160eps7}
\end{figure}

Our canonical basis allows us to straightforwardly evaluate all master integrals numerically in the whole kinematic space. One option is to repeat the procedure of \cref{sec:series and boundaries} at sufficiently many expansion points, and to match the resulting series expansions. Here, we instead use the numerical solver provided with~\cite{Czakon:2026tog}. In this subsection, we use the variable $x = 1/z = p^2/m^2$, in which the singular points of the differential equation, see \cref{sec:analyticresults}, are located at
\begin{equation}
  x\in\{-7,-3,-1,0,\tfrac19,1,2,9,25,\infty\}\,.
\end{equation}
We use the canonical basis with the special functions defined around $x=1$, see \cref{sec:onshell}, and numerical boundary values at $x_0=1.0625$, just above the threshold at $x=1$. This is necessary, as the numerical evolution code cannot
start from a divergent boundary point for the connection matrix. 
From there, the system is integrated along a straight path, deformed into the upper half of the complex plane to select the $x+i0$ prescription. As an illustration, we show the real and imaginary parts of the weight-$7$ contribution to the canonical top-sector integral, $J_{160}^{(7)}$, for $x\in [-10,30]$ in \cref{fig:J160eps7}. We pick this function as a representative since it is the top sector integral, which couples to the subsectors in the differential equation, at the maximal weight we computed.
We stress here that the whole system is integrated at once, which means that the 
full set of integrals through weight seven can be plotted in the same way.

\begin{table}[t]
\centering
\begin{tabular}{rcccc}
\toprule
$x$ & region & crossed & time [s] & error $\Delta$ \\
\midrule
$1.25$  & $(1,2)$          & 0 & 0.06 & $6\cdot10^{-10}$ \\
$1.9$   & $(1,2)$          & 0 & 0.10 & $1\cdot10^{-9}$ \\
$3$     & $(2,9)$          & 1 & 0.10 & $2\cdot10^{-9}$ \\
$5$     & $(2,9)$          & 1 & 0.12 & $2\cdot10^{-8}$ \\
$11$    & $(9,25)$         & 2 & 0.17 & $1\cdot10^{-8}$ \\
$20$    & $(9,25)$        & 2 & 0.17 & $3\cdot10^{-8}$ \\
$30$    & $(25,\infty)$    & 3 & 0.19 & $6\cdot10^{-8}$ \\
$1000$  & $(25,\infty)$    & 3 & 0.28 & $2\cdot10^{-7}$ \\
\midrule
$0.5$   & $(1/9,1)$        & 1 & 0.11 & $6\cdot10^{-7}$ \\
$0.05$  & $(0,1/9)$        & 2 & 0.19 & $2\cdot10^{-6}$ \\
$-0.5$  & $(-1,0)$         & 3 & 0.17 & $1\cdot10^{-6}$ \\
$-2$    & $(-3,-1)$        & 4 & 0.15 & $2\cdot10^{-7}$ \\
$-5$    & $(-7,-3)$        & 5 & 0.18 & $2\cdot10^{-7}$ \\
$-9$    & $(-\infty,-7)$   & 6 & 0.27 & $2\cdot10^{-7}$ \\
$-100$  & $(-\infty,-7)$   & 6 & 0.28 & $7\cdot10^{-7}$ \\
\bottomrule
\end{tabular}
\caption{Evaluation time and accuracy of our numerical results for the canonical master integrals at points in different regions, using the numerical solver of~\cite{Czakon:2026tog}.
The column ``error'' gives the maximum deviation, over all 160 integrals and $\varepsilon^1,\dots,\varepsilon^7$, of the final result from a reference evaluation in double-double precision (about 32 digits) with tolerance $\tau = 10^{-20}$. ``Crossed'' refers to the number of singular points between the boundary point $x_0$ and the evaluation point $x$.}\label{tab:regions}
\end{table}

In \cref{tab:regions}, we list evaluation times and errors for points in all regions bounded by the singular points. 
To be precise, we integrate along a path deformed into the upper half-plane with deformation parameter $0.25$ (the path reaches a maximal imaginary part of $0.25\,|x-x_0|$), and an integrator tolerance $\tau = 10^{-10}$. 
The tolerance controls the step size: a step is accepted if $\max_i |\delta f_i|/(|f_i|+1) < \tau$, where $f_i$ runs over all canonical integrals at $\varepsilon^1$ to $\varepsilon^7$ and the auxiliary functions, and $\delta f_i$ is the integrator's internal estimate of the error made in that step. The error shown in \cref{tab:regions} is the maximum deviation across all 160 master integrals and orders $\varepsilon^1$ to $\varepsilon^7$ from a reference evaluation in double-double precision, with tolerance $10^{-20}$, along the same contour.
The reference is accurate to better than $10^{-16}$, as we checked by comparing with tolerance $10^{-18}$, so the error is that of the double-precision evaluation itself.

In all cases, a single evaluation takes well below one second. As expected, the evaluation time grows with the number of singular points between $x_0$ and $x$, from about $0.06$\,s in the boundary region $(1,2)$ to about $0.3$\,s when all six singular points below the threshold are passed. On the other hand, the distance to the evaluation point plays only a minor role. More precisely, the time is governed by how closely the path passes the singular points, since the adaptive step size has to shrink proportionally to the distance from a singularity. A moderate deformation of the contour ($\approx 0.25$--$0.5$) turns out to be close to optimal.

The accuracy is limited by double-precision round-off rather than by the integrator tolerance. 
Below $\tau \approx 10^{-8}$, the error $\Delta$ no longer improves. It is about $10^{-9}$ near the boundary point, in agreement with an independent series solution at $x=1.25$, $10^{-9}$--$10^{-7}$ for $x>2$, and $10^{-7}$--$10^{-6}$ below the threshold. The main loss of precision occurs once, when the path passes around the threshold at $x=1$, only $1/16$ away from the boundary point, and does not grow with the number of further singular points crossed. At $x=1000$, the slightly larger error is due to the growth of the integrals themselves, and is consistent with a comparison against \texttt{AMFlow}, which shows deviations of order $10^{-7}$. Note that there are several ways to get higher precision or improve the evaluation times further, if required. First, the solver can be run in higher precision if necessary, as we have done here to determine the error of the double precision run.  
Also, one could first generate a single point in every region with higher precision. 
These can then be used as new boundary points for all further evaluations, such that they never need to cross any singular point.
We do not elaborate on this further here, and postpone a more detailed numerical study to the case of the
full electron self-energy.

\section{Conclusions}\label{sec:conclusion}

In this work, we have analytically computed the four-loop ladder integral family contributing to the electron self-energy in QED, with finite electron mass $m$ and a general off-shell external momentum $p$.
Importantly, this single family couples polylogarithmic, elliptic, and Calabi-Yau sectors, making this an interesting problem for studying non-trivial couplings between these geometries. 
%Moreover, its on-shell values span all non-polylogarithmic constants in the planar contribution to the four-loop QED anomalous magnetic moment $(g-2)$ of the electron.

In total, the family contains 160 master integrals, which we calculate using the method of differential equations.
To construct a canonical basis, we take suitable linear combinations of Feynman integrals such that their leading singularities are constant. This requires us to introduce square roots, periods of the involved geometries, and iterated integrals over them.
We present details of our calculation for four exemplary sectors, illustrating the construction for several geometries and their couplings to each other.

From the canonical basis, we obtain the general analytic solution to the differential equation system in terms of iterated integrals over kernels that are either rational, algebraic, or of elliptic, Calabi-Yau, or mixed type. Notably, all extra special functions appearing in the kernels are themselves (iterative) solutions of differential equations with rational coefficients. Using this structure, we derive local series solutions of the differential equation system around two special points, that are the MUM point $p^2/m^2 = \infty$, where the analytic structure is the simplest, and the on-shell point $p^2=m^2$.
While the boundary values at the MUM point are given purely by a specific type of polylogarithmic numbers, the structure in the on-shell limit is much richer and involves constants of Calabi-Yau type. We find analytical representations for the polylogarithmic constants and observe that all non-polylogarithmic ones coincide with those appearing in the planar contribution to the four-loop QED anomalous magnetic moment $(g-2)$ of the electron~\cite{Laporta:2017okg}.

Finally, we illustrate how the canonical master integrals can be evaluated conveniently using the numerical solver provided with~\cite{Czakon:2026tog}, and we study how the runtime and error depend on the distance and number of singularities between the boundary and evaluation point.

\acknowledgments

We thank Claude Duhr and Benjamin Sauer for useful discussions and collaboration on related projects.
This work was partly funded by the European Union through the European Research Council under the grant agreements 949279 (ERC Starting Grant HighPHun (LT, CN)), 101167287 (ERC Synergy Grant  MaScAmp (CN)) and by the Deutsche Forschungsgemeinschaft (DFG, German Research Foundation) through the Excellence Cluster ORIGINS EXC-2094-390783311 (LT) and the research unit FOR 5582 – Projektnummer 508889767 (FF).
Views and opinions expressed are however those of the author(s) only and do not necessarily reflect those of the European Union or the European Research Council. Neither the European Union nor the granting authority can be held responsible for them.

\appendix

\crefalias{section}{appendix}

\crefname{appendix}{appendix}{appendices}
\Crefname{appendix}{Appendix}{Appendices}

\newpage
\section{Graphs for irreducible sectors}
\label{sec:allgraphs}
In this appendix, we list the graphs for all master integrals calculated in this paper, where upper numbers are the sector IDs and lower ones give the number of masters per sector. We indicate the geometry of the maximal cut on the top right, unless it is polylogarithmic.

\begin{center}
\includegraphics[width=0.93\linewidth]{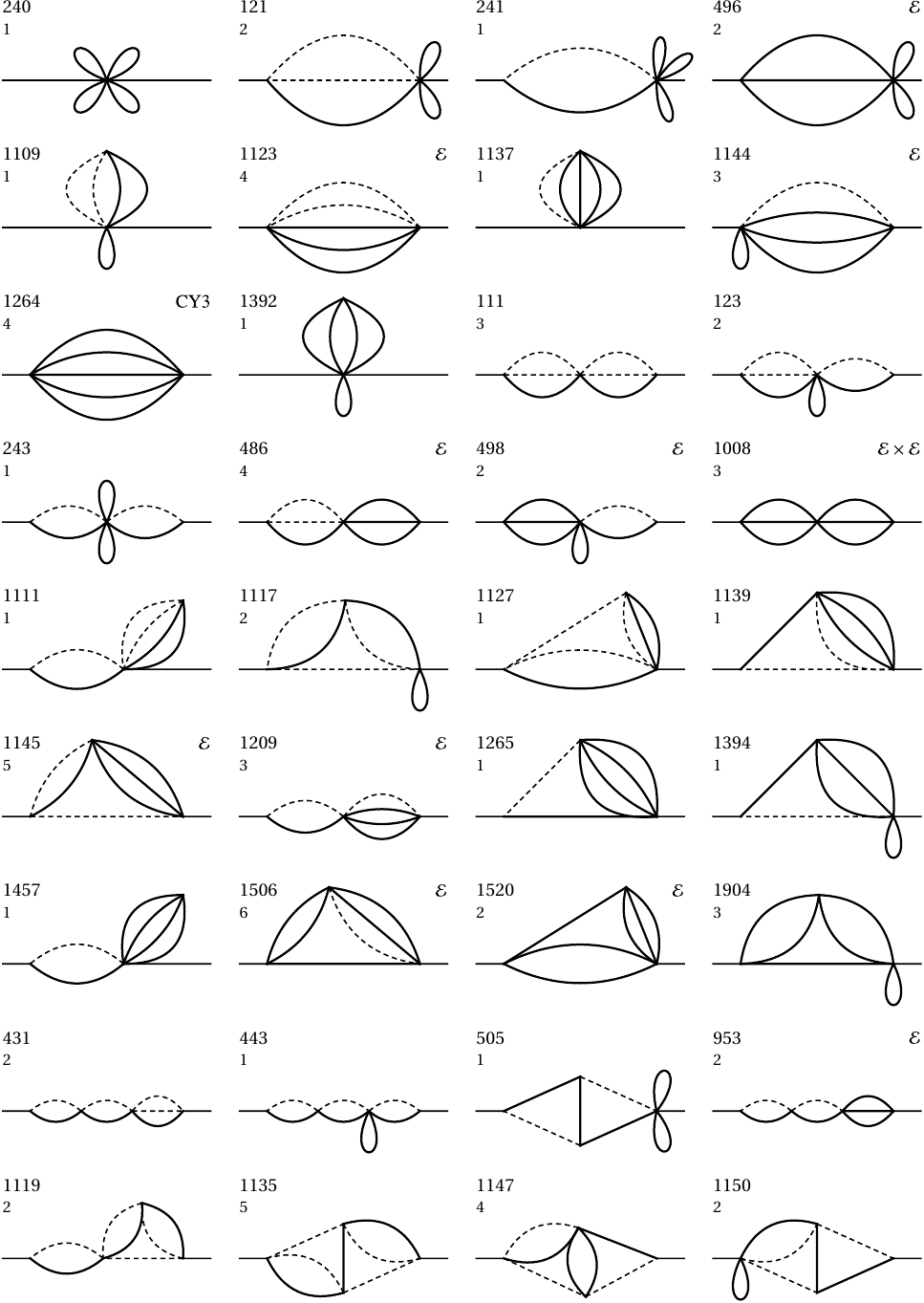}
\end{center}

\begin{center}
\includegraphics[width=0.93\linewidth]{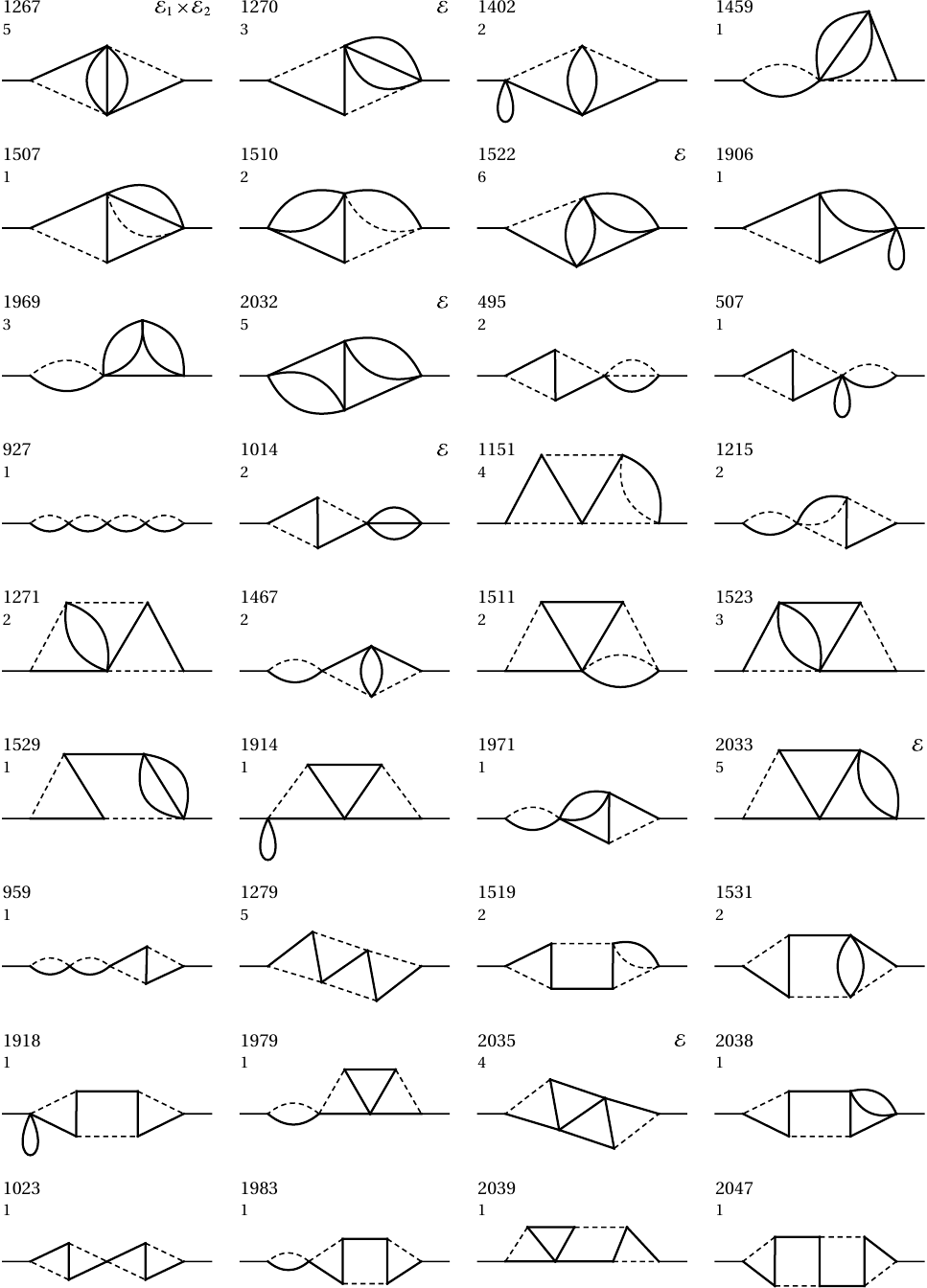}
\end{center}
\newpage

\section{Pre-canonical basis}
\label{app:precan}

Here we give our pre-canonical basis, which consists of linear combinations of Feynman integrals with rational coefficients, meaning that no special functions have been introduced yet. 
While we mostly follow the notation introduced in \cref{sec:general}, for brevity we sometimes write $I_{a_1 \dots a_{14}}\big[N\big]$ to denote the same integral with the polynomial $N$ in the loop momenta inserted in the numerator. 
Of course, one can express these polynomials in terms of the complete set of propagators \cref{eq:intfam}, but doing so would only make the subsequent expressions longer.
Furthermore, $G(q_1,q_2,q_3)=\det(q_i\cdot q_j)$ for $i,j=1,2,3$, is the Gram determinant. 
$D^{-}$ denotes the dimension shift $d\to d-2$, i.e., $d=2-2\epsilon$ instead of $4-2\epsilon$. 
Moreover, we set $m^2=1$ and use the dimensionless variable $z=m^2/p^2$. 
With this, we define the following pre-canonical basis:
\allowdisplaybreaks
\begin{align*}
P_{1}^{(240)} &= D^{-}I_{00001111000} \\[4pt]
P_{2}^{(121)} &= D^{-}I_{10011110000} \\[4pt]
P_{3}^{(121)} &= D^{-}I_{10011110(-1)00} \\[4pt]
P_{4}^{(241)} &= D^{-}I_{10001111000} \\[4pt]
P_{5}^{(496)} &= D^{-}I_{00001111100} \\[4pt]
P_{6}^{(496)} &= D^{-}I_{00001121100} \\[4pt]
P_{7}^{(1109)} &= D^{-}I_{10101010(-1)01} \\[4pt]
P_{8}^{(1123)} &= D^{-}I_{11000110001} \\[4pt]
P_{9}^{(1123)} &= D^{-}I_{11000110002} \\[4pt]
P_{10}^{(1123)} &= D^{-}I_{11000110001}\Big[-\frac{3 z+1}{3 z} + (k_1+k_2-k_3-k_4-p)^2\Big] \\[4pt]
P_{11}^{(1123)} &= D^{-}I_{11000110001}\big[(k_1+k_2-p)^2\big] \\[4pt]
P_{12}^{(1137)} &= D^{-}I_{10001110001} \\[4pt]
P_{13}^{(1144)} &= D^{-}I_{00011110001} \\[4pt]
P_{14}^{(1144)} &= D^{-}I_{00011110002} \\[4pt]
P_{15}^{(1144)} &= D^{-}I_{00011110001}\Big[-\frac{3 z+1}{3 z} + (k_1+k_2-k_3-k_4+p)^2\Big] \\[4pt]
P_{16}^{(1264)} &= D^{-}I_{00001111001} \\[4pt]
P_{17}^{(1264)} &= D^{-}I_{00002111001} \\[4pt]
P_{18}^{(1264)} &= D^{-}I_{00003111001} \\[4pt]
P_{19}^{(1264)} &= D^{-}I_{00004111001} \\[4pt]
P_{20}^{(1392)} &= D^{-}I_{00001110101} \\[4pt]
P_{21}^{(111)} &= D^{-}I_{11110110000} \\[4pt]
P_{22}^{(111)} &= D^{-}I_{11110110(-1)00} \\[4pt]
P_{23}^{(111)} &= D^{-}I_{1111(-1)110(-1)00} \\[4pt]
P_{24}^{(123)} &= D^{-}I_{11011110000} \\[4pt]
P_{25}^{(123)} &= D^{-}I_{11011110(-1)00} \\[4pt]
P_{26}^{(243)} &= D^{-}I_{11001111000} \\[4pt]
P_{27}^{(486)} &= D^{-}I_{01100111100} \\[4pt]
P_{28}^{(486)} &= D^{-}I_{01100121100} \\[4pt]
P_{29}^{(486)} &= D^{-}I_{011001111(-1)0} \\[4pt]
P_{30}^{(486)} &= D^{-}I_{011001211(-1)0} \\[4pt]
P_{31}^{(498)} &= D^{-}I_{01001111100} \\[4pt]
P_{32}^{(498)} &= D^{-}I_{01001121100} \\[4pt]
P_{33}^{(1008)} &= D^{-}I_{00001111110} \\[4pt]
P_{34}^{(1008)} &= D^{-}I_{00002111110} \\[4pt]
P_{35}^{(1008)} &= D^{-}I_{00002121110} \\[4pt]
P_{36}^{(1111)} &= D^{-}I_{11101010(-1)01} \\[4pt]
P_{37}^{(1117)} &= -D^{-}I_{10111010(-1)01} \\[4pt]
P_{38}^{(1117)} &= D^{-}I_{10111010(-2)01} \\[4pt]
P_{39}^{(1127)} &= D^{-}I_{11100110(-1)01} \\[4pt]
P_{40}^{(1139)} &= D^{-}I_{11001110001} \\[4pt]
P_{41}^{(1145)} &= D^{-}I_{10011110(-1)01} \\[4pt]
P_{42}^{(1145)} &= D^{-}I_{10012110(-1)01} \\[4pt]
P_{43}^{(1145)} &= D^{-}I_{10011110001} \\[4pt]
P_{44}^{(1145)} &= D^{-}I_{1001111(-1)001} \\[4pt]
P_{45}^{(1145)} &= D^{-}I_{1001111(-1)(-1)01} \\[4pt]
P_{46}^{(1209)} &= D^{-}I_{10011101001} \\[4pt]
P_{47}^{(1209)} &= D^{-}I_{10012101001} \\[4pt]
P_{48}^{(1209)} &= D^{-}I_{10011101(-1)01} \\[4pt]
P_{49}^{(1265)} &= D^{-}I_{10001111001} \\[4pt]
P_{50}^{(1394)} &= D^{-}I_{01001110101} \\[4pt]
P_{51}^{(1457)} &= D^{-}I_{10001101101} \\[4pt]
P_{52}^{(1506)} &= -\frac{1}{2}\,D^{-}I_{010001111(-1)1} \\[4pt]
P_{53}^{(1506)} &= D^{-}I_{01000111102} \\[4pt]
P_{54}^{(1506)} &= -\frac{1}{2}\,D^{-}I_{010(-1)01111(-1)1} \\[4pt]
P_{55}^{(1506)} &= -4\,D^{-}I_{010(-1)0111101} + D^{-}I_{010(-2)0111101} \\[4pt]
P_{56}^{(1506)} &= -4\,D^{-}I_{01000111101} + D^{-}I_{010(-1)0111101} \\[4pt]
P_{57}^{(1506)} &= D^{-}I_{010(-1)0111101} \\[4pt]
P_{58}^{(1520)} &= D^{-}I_{00001111101} \\[4pt]
P_{59}^{(1520)} &= D^{-}I_{00001121101} \\[4pt]
P_{60}^{(1904)} &= D^{-}I_{00001110111} \\[4pt]
P_{61}^{(1904)} &= D^{-}I_{00(-1)01110111} \\[4pt]
P_{62}^{(1904)} &= D^{-}I_{00(-2)01110111} \\[4pt]
P_{63}^{(431)} &= D^{-}I_{11110101100} \\[4pt]
P_{64}^{(431)} &= D^{-}I_{1111(-1)101100} \\[4pt]
P_{65}^{(443)} &= D^{-}I_{11011101100} \\[4pt]
P_{66}^{(505)} &= I_{10011111100} \\[4pt]
P_{67}^{(953)} &= D^{-}I_{10011101110} \\[4pt]
P_{68}^{(953)} &= D^{-}I_{10012101110} \\[4pt]
P_{69}^{(1119)} &= D^{-}I_{11111010(-2)01} \\[4pt]
P_{70}^{(1119)} &= \frac{z-1}{z}\,D^{-}I_{11111010(-1)01} + D^{-}I_{11111010(-2)01} \\[4pt]
P_{71}^{(1135)} &= D^{-}I_{11110110001}\Big[-\frac{z-1}{z^2} + \frac{(k_2-k_4)^2 (k_1-k_3-p)^2 (z-1)}{z}\Big] \\[4pt]
P_{72}^{(1135)} &= D^{-}I_{11110110001}\big[(k_1-k_3-p)^2\big((k_2-k_4)^4-(k_2-k_4)^2-p^2\big)+p^2\big] \\[4pt]
P_{73}^{(1135)} &= \frac{(z-1)^2}{z^2}\,D^{-}I_{11110110001} + \frac{2 (z-1)}{z}\,D^{-}I_{111101100(-1)1} \\[4pt]
P_{74}^{(1135)} &= 3\,D^{-}I_{11110110(-1)01} - \frac{2 z+1}{z}\,D^{-}I_{111101100(-1)1} \\
  &\quad + D^{-}I_{11110110(-2)01} + D^{-}I_{11110110(-1)(-1)1} \\[4pt]
P_{75}^{(1135)} &= D^{-}I_{11110110001}\big[\big((k_2-k_4)^2\big)^2 (k_1-k_3-p)^2\big] \\[4pt]
P_{76}^{(1147)} &= -\frac{(z-1) (z+1)}{z^2}\,D^{-}I_{11011110001}\big[(k_1+k_2-k_4)^2\big] \\[4pt]
P_{77}^{(1147)} &= \frac{z-1}{z}\,D^{-}I_{11011110001}\big[\big((k_2-k_4)^2-1\big)\,(k_1+k_2-k_4)^2\big] \\[4pt]
P_{78}^{(1147)} &= \frac{1}{z}\,D^{-}I_{11011110001}\big[(k_1+k_2-k_4)^2\big((k_1+k_2-k_4)^2+(z-1)\big((k_2-k_4)^2-p^2\big)\big)\big] \\[4pt]
P_{79}^{(1147)} &= D^{-}I_{11011110001}\big[(k_1-k_4)^2 (k_1+k_2-k_4)^2\big] \\[4pt]
P_{80}^{(1150)} &= -\frac{z-1}{z}\,D^{-}I_{01111110001}\big[k_1^2-k_3^2+(k_1-k_3-p)^2-p^2\big] \\[4pt]
P_{81}^{(1150)} &= -4\,D^{-}I_{01111110001}\big[G(k_1,k_3,p)\big] \\[4pt]
P_{82}^{(1267)} &= D^{-}I_{11001111001}\Big[-\frac{(k_1+k_2)^2}{z} + k_1^2 (k_2+p)^2\Big] \\[4pt]
P_{83}^{(1267)} &= \partial_z P_{82} \\[4pt]
P_{84}^{(1267)} &= 2\,D^{-}I_{11001111001}\big[k_2\cdot (k_1-p)\big] \\[4pt]
P_{85}^{(1267)} &= \partial_z P_{84} \\[4pt]
P_{86}^{(1267)} &= 4\,D^{-}I_{11001111001}\big[G(k_1,k_2,p)\big] \\[4pt]
P_{87}^{(1270)} &= D^{-}I_{01101111001}\big[p^2-k_1^2+k_3^2-(k_1-k_3-p)^2\big] \\[4pt]
P_{88}^{(1270)} &= \partial_z P_{87} \\[4pt]
P_{89}^{(1270)} &= 2\,D^{-}I_{01101111001}\big[(k_1-k_3)\cdot(k_1-p)\,\big(k_1^2(k_1-k_3-p)^2-p^2k_3^2\big)\big] \\[4pt]
P_{90}^{(1402)} &= -\frac{(z-1) (z+1)}{z^2}\,D^{-}I_{01011110101}\big[(k_1+k_2-k_4)^2\big] \\[4pt]
P_{91}^{(1402)} &= D^{-}I_{01011110101}\Big[\frac{(k_1+k_2-k_4)^2 (z-1)^2}{z^2} + \frac{\big((k_1+k_2-k_4)^2\big)^2}{z}\Big] \\[4pt]
P_{92}^{(1459)} &= \frac{(z-1)^2}{z^2}\,D^{-}I_{11001101101} \\[4pt]
P_{93}^{(1507)} &= \frac{z-1}{z^2}\big[-(z+1)\,D^{-}I_{10011101001}+2z\,D^{-}I_{11000110101}+2(z-1)\,D^{-}I_{11000111101}\big] \\[4pt]
P_{94}^{(1510)} &= D^{-}I_{01100111101}\Big[\big(p^2-(k_1-k_3-p)^2\big) \\
  &\quad \times\big(\tfrac{z}{1-z}\big((k_2-k_4)^2-p^2\big)^2+\big[(k_2-k_4)^2-m^2\big](k_2-k_4+p)^2\big)\Big] \\[4pt]
P_{95}^{(1510)} &= 4\,D^{-}I_{01100111101}\big[G(k_1-k_3,k_2-k_4,p)\big] \\[4pt]
P_{96}^{(1522)} &= D^{-}I_{010(-2)1111101} \\[4pt]
P_{97}^{(1522)} &= D^{-}I_{01001111101}\big[(k_1+k_2-k_4)^2 (k_2-k_4+p)^2\big] \\[4pt]
P_{98}^{(1522)} &= D^{-}I_{010(-1)1111101} \\[4pt]
P_{99}^{(1522)} &= D^{-}I_{01001111101}\big[\big((k_1+k_2-k_4)^2\big)^2\big] \\[4pt]
P_{100}^{(1522)} &= D^{-}I_{01001111101}\big[(k_1+k_2-k_4)^2\big] \\[4pt]
P_{101}^{(1522)} &= \partial_z\big[2\,P_{98} -\frac{z+1}{z}\,P_{100}\big] \\[4pt]
P_{102}^{(1906)} &= I_{01001110211} \\[4pt]
P_{103}^{(1969)} &= D^{-}I_{10001101111} \\[4pt]
P_{104}^{(1969)} &= D^{-}I_{10(-1)01101111} \\[4pt]
P_{105}^{(1969)} &= D^{-}I_{10(-2)01101111} \\[4pt]
P_{106}^{(2032)} &= \frac{2 (z-1)}{z}\,D^{-}I_{00001111111} + D^{-}I_{00(-1)(-1)1111111} \\[4pt]
P_{107}^{(2032)} &= \partial_z P_{106} \\[4pt]
P_{108}^{(2032)} &= -\frac{3 (z-1)}{z}\,D^{-}I_{00001111111} + \frac{z-1}{z}\,D^{-}I_{00(-1)01111111} \\
  &\quad + \frac{z-1}{z}\,D^{-}I_{000(-1)1111111} + \frac{z+1}{4 z}\,D^{-}I_{00(-1)(-1)1111111} \\[4pt]
P_{109}^{(2032)} &= -\frac{2 (z-1)}{z}\,D^{-}I_{00001111111} + \frac{z-1}{z}\,D^{-}I_{000(-1)1111111}  + D^{-}I_{00(-1)(-1)1111111}  \\[4pt]
P_{110}^{(2032)} &= I_{00002121111} \\[4pt]
P_{111}^{(495)} &= -4\,D^{-}I_{11110111100}\big[G(k_2,k_4,p)\big] \\[4pt]
P_{112}^{(495)} &= -4\,D^{-}I_{1111(-1)111100}\big[G(k_2,k_4,p)\big] \\[4pt]
P_{113}^{(507)} &= -4\,D^{-}I_{11011111100}\big[G(k_2,k_4,p)\big] \\[4pt]
P_{114}^{(927)} &= D^{-}I_{11111001110} \\[4pt]
P_{115}^{(1014)} &= -4\,D^{-}I_{01101111110}\big[G(k_1,k_3,p)\big] \\[4pt]
P_{116}^{(1014)} &= \partial_z P_{115} \\[4pt]
P_{117}^{(1151)} &= -\frac{4 \epsilon^2 (2 \epsilon-1) \left(z^2+3\right)}{(z-3) (z-1) z}\,I_{11111120001}  - \frac{2 \epsilon (2 \epsilon-1) \left(z^2+z+4\right)}{(z-3) (z-1) z}\,I_{11111120002} \\
  &\quad - \frac{2 (2 \epsilon-1)}{(z-3) (z-1)}\,I_{11111220001}  - \frac{2 \epsilon (2 \epsilon-1)}{(z-3) z^2}\,I_{11121120001}  \\[4pt]
P_{118}^{(1151)} &= -\frac{12 \epsilon^2 (2 \epsilon-1)}{z}\,I_{11111120001} - \frac{8 \epsilon (2 \epsilon-1)}{z}\,I_{11111120002}  + \frac{2 \epsilon (2 \epsilon-1)}{z^2}\,I_{11121120001}  \\[4pt]
P_{119}^{(1151)} &= \frac{\epsilon (2 \epsilon-1)}{2 z^2}\,I_{11121120001} \\[4pt]
P_{120}^{(1151)} &= \frac{\epsilon^2 (2 \epsilon-1)}{z}\,I_{11111120001} \\[4pt]
P_{121}^{(1215)} &= \frac{(z-1)^2}{z^2}\,D^{-}I_{11111101001}\big[k_1^2-k_3^2+(k_1-k_3-p)^2-p^2\big] \\[4pt]
P_{122}^{(1215)} &= \frac{4 (z-1)}{z}\,D^{-}I_{11111101001}\big[G(k_1,k_3,p)\big] \\[4pt]
P_{123}^{(1271)} &= -\frac{2 \epsilon (2 \epsilon-1) (z-1)}{z^2}\,I_{12101111002} \\[4pt]
P_{124}^{(1271)} &= \frac{\epsilon^2 (2 \epsilon-1)}{z}\,I_{11101111002} \\[4pt]
P_{125}^{(1467)} &= \frac{(z-1)^2 (z+1)}{z^3}\,D^{-}I_{11011101101}\big[(k_1+k_2-k_4)^2\big] \\[4pt]
P_{126}^{(1467)} &= -\frac{z-1}{z^2}\,D^{-}I_{11011101101}\big[(k_1+k_2-k_4)^2\big((k_1+k_2-k_4)^2+\tfrac{(z-1)^2}{z}\big)\big] \\[4pt]
P_{127}^{(1511)} &= -\frac{2 \epsilon^2 (2 \epsilon-1)}{z}\,I_{11100211101} \\[4pt]
P_{128}^{(1511)} &= \frac{2 \epsilon (2 \epsilon-1) (z-1)}{z^2}\,I_{12100211101} \\[4pt]
P_{129}^{(1523)} &= I_{11001211101} \\[4pt]
P_{130}^{(1523)} &= I_{12001211101} \\[4pt]
P_{131}^{(1523)} &= I_{11002211101} \\[4pt]
P_{132}^{(1529)} &= D^{-}I_{10011111101}\big[G(k_2,k_4,p)\big] \\[4pt]
P_{133}^{(1914)} &= \frac{1}{z}\,I_{01011120111} \\[4pt]
P_{134}^{(1971)} &= -\frac{z-1}{z}\,D^{-}I_{11001101111}\big[G(k_1,k_3,p)\big] \\[4pt]
P_{135}^{(2033)} &= \frac{1}{z}\,I_{10001211121} \\[4pt]
P_{136}^{(2033)} &= \partial_z P_{135} \\[4pt]
P_{137}^{(2033)} &= \frac{1}{z}\,I_{10(-1)01211121} \\[4pt]
P_{138}^{(2033)} &= \frac{1}{z}\,I_{10001211111} \\[4pt]
P_{139}^{(2033)} &= \frac{1}{z}\,I_{10001211112} \\[4pt]
P_{140}^{(959)} &= -\frac{4 (z-1)^2}{z^2}\,D^{-}I_{11111101110}\big[G(k_1,k_3,p)\big] \\[4pt]
P_{141}^{(1279)} &= -\frac{\epsilon^2 (2 \epsilon-1)}{2 z}\,I_{11111111002} - \frac{\epsilon (2 \epsilon-1)}{2 z (z+1)}\,I_{11111121001}  - \frac{\epsilon^2 (2 \epsilon-1) (z-1)}{2 z^2 (z+1)}\,I_{21111111001}  \\[4pt]
P_{142}^{(1279)} &= \frac{\epsilon(2\epsilon-1)}{z^2}\Big[\epsilon(3z+1)\,I_{11111111002}+\frac{4z}{z+1}\,I_{11111121001}+\frac{4\epsilon(z-1)}{z+1}\,I_{21111111001}\Big] \\[4pt]
P_{143}^{(1279)} &= -\frac{8 \epsilon (2 \epsilon-1) (z-1)^2}{z^3}\,I_{22111111001} \\[4pt]
P_{144}^{(1279)} &= -\frac{4 \epsilon^2 (2 \epsilon-1) (z-1)}{z^2}\,I_{21111111001} \\[4pt]
P_{145}^{(1279)} &= -\frac{\epsilon^3 (2 \epsilon-1)}{2 z}\,I_{11111111001} \\[4pt]
P_{146}^{(1519)} &= -\frac{z-1}{z^2}\,I_{11110211101} \\[4pt]
P_{147}^{(1519)} &= \frac{(z-1)^2}{z^2 (z+1)}\,I_{11110211102} \\[4pt]
P_{148}^{(1531)} &= -\frac{z-1}{z^2}\,I_{11011211101} \\[4pt]
P_{149}^{(1531)} &= \frac{(z-1) (z+1)}{z^3}\,I_{11012211101} \\[4pt]
P_{150}^{(1918)} &= I_{01111120111} \\[4pt]
P_{151}^{(1979)} &= I_{11011102111} \\[4pt]
P_{152}^{(2035)} &= I_{11001111112} \\[4pt]
P_{153}^{(2035)} &= \partial_z P_{152} \\[4pt]
P_{154}^{(2035)} &= I_{110(-1)1111112} \\[4pt]
P_{155}^{(2035)} &= I_{11001111111} \\[4pt]
P_{156}^{(2038)} &= I_{01101121111} \\[4pt]
P_{157}^{(1023)} &= I_{11111111110} \\[4pt]
P_{158}^{(1983)} &= I_{11111102111} \\[4pt]
P_{159}^{(2039)} &= I_{11101111111} \\[4pt]
P_{160}^{(2047)} &= I_{11111111111}
\end{align*}

\section{Special functions for our canonical basis}
\label{app:Gfunctions}

In this appendix, we collect the square roots and the defining differential equations of the transcendental functions $G_1,\dots,G_{27}$ needed for our canonical basis, see \cref{sec:can basis}. The periods $\pi_0$ $\varpi_0$, $\varpi_0^{\text{(shift)}}, \varpi_0^{\text{(inv)}}$ and the structure series $\alpha$ are defined in \cref{sec:general}.

\begin{align*}
r_1 &= \sqrt{1-z}\,, & r_2 &= \sqrt{1-9z}\,, & r_3 &= \sqrt{1+3z}\,, & r_4 &= \sqrt{1+z} \,.
\end{align*}

\begin{align*}
G_{1}'(z) &= \frac{\pi_0(z)^2 \left(191 z^2+2 z-1\right) \left(48825 z^4-11684 z^3-234 z^2-36 z-7\right)}{\alpha(z)z^2 (z-1)^2 (9 z-1)^2 (25 z-1)^2} \\[2pt]
G_{2}'(z) &= -\frac{G_{1}(z)^2 \alpha(z)^2 z}{\pi_0(z)^2 (z-1) (9 z-1) (25 z-1)} + \frac{q_{1}(z) \pi_0(z)^2}{4z^3 (z-1)^3  (9 z-1)^3 (25 z-1)^3} \\
  &\quad - \frac{q_{2}(z) \pi_0'(z)^2}{2z (z-1)  (9 z-1) (25 z-1)} \\[2pt]
G_{3}'(z) &= -\frac{G_{1}(z) \alpha(z)^2 z}{\pi_0(z)^2 (z-1) (9 z-1) (25 z-1)} \\[2pt]
G_{4}'(z) &= \frac{G_{2}(z)}{\alpha(z) z} + \frac{q_{2}(z) \pi_0(z) \pi_0'(z)}{2 \alpha(z)z^2 (z-1)  (9 z-1) (25 z-1)} \\[2pt]
G_{5}'(z) &= \frac{2 \pi_0(z) (5 z+1)}{5 z(z-1)^2 } \\[2pt]
G_{6}'(z) &= \frac{3 \pi_0(z) \varpi_0(z) \left(459 z^4+54 z^3+52 z^2-54 z+1\right)}{10 z^2 (z-1)^2 (9 z-1)^2} + \frac{\varpi_0(z) \pi_0'(z) \left(-99 z^2+10 z-7\right)}{5 z(z-1)  (9 z-1)} \\[2pt]
G_{7}'(z) &= -\frac{G_{6}(z) z}{\varpi_0(z)^2 (z-1) (9 z-1)} \\[2pt]
G_{8}'(z) &= -\frac{G_{6}(z)}{\alpha(z) z} + \frac{\pi_0(z) \varpi_0(z) \left(-99 z^2+10 z-7\right)}{5 \alpha(z) z^2(z-1)  (9 z-1)} \\[2pt]
G_{9}'(z) &= \frac{G_{3}(z) G_{6}(z)}{\alpha(z) z} + \frac{G_{3}(z) \pi_0(z) \varpi_0(z) \left(99 z^2-10 z+7\right)}{5 \alpha(z)z^2 (z-1)  (9 z-1)} + \frac{G_{6}(z)}{z (25 z-1)} \\
  &\quad + \frac{2 \pi_0(z) \varpi_0(z) (3 z-1) \left(675 z^4+3459 z^3+461 z^2+17 z-4\right)}{5 z^2(z-1)^2  (9 z-1)^2 (25 z-1)} \\[2pt]
G_{10}'(z) &= -\frac{\varpi_0^{\text{(shift)}}(z) \varpi_0^{\text{(inv)}}(z)}{z} + 2 \varpi_0^{\text{(inv)}}(z) \big(\varpi_0^{\text{(shift)}}\big)'(z) (7 z+5) \\[2pt]
G_{11}'(z) &= \frac{4 \pi_0(z) \varpi_0^{\text{(shift)}}(z) (2 z+1)}{5 z^3} - \frac{2 \varpi_0^{\text{(shift)}}(z) \pi_0'(z) \left(70 z^3+75 z^2-z+2\right)}{5 z^2} \\
  &\quad - \frac{2 \varpi_0^{\text{(shift)}}(z) \pi_0''(z) \left(35 z^3+103 z^2-16 z-2\right)}{5 z} \\[2pt]
G_{12}'(z) &= -\frac{4 \pi_0(z) \varpi_0^{\text{(inv)}}(z)}{5 z} - \frac{2}{5} \varpi_0^{\text{(inv)}}(z) \pi_0'(z) (55 z+43) - 2 \varpi_0^{\text{(inv)}}(z) \pi_0''(z) z (11 z+13) \\[2pt]
G_{13}'(z) &= \frac{G_{11}(z) z}{\big(\varpi_0^{\text{(shift)}}(z)\big)^2 (z-1) (2 z-1) (7 z+1)} \\[2pt]
G_{14}'(z) &= \frac{G_{12}(z)}{\big(\varpi_0^{\text{(inv)}}(z)\big)^2 z(z-9) (z-1) } \\[2pt]
G_{15}'(z) &= -\frac{G_{11}(z)}{\alpha(z) z} - \frac{2 \pi_0(z) \varpi_0^{\text{(shift)}}(z) \left(35 z^3-28 z^2+15 z+4\right)}{5 \alpha(z) z^3} \\
  &\quad - \frac{4 \varpi_0^{\text{(shift)}}(z) \pi_0'(z) \left(35 z^3+103 z^2-16 z-2\right)}{5 \alpha(z) z^2} \\
  &\quad + \frac{2 \pi_0(z) \varpi_0^{\text{(shift)}}(z) \alpha'(z) \left(35 z^3+103 z^2-16 z-2\right)}{5 \alpha(z)^2 z^2} \\[2pt]
G_{16}'(z) &= -\frac{G_{12}(z)}{\alpha(z) z} + \frac{44 \pi_0(z) \varpi_0^{\text{(inv)}}(z)}{5 \alpha(z) z} - \frac{4 \varpi_0^{\text{(inv)}}(z) \pi_0'(z) (11 z+13)}{\alpha(z)} \\
  &\quad + \frac{2 \pi_0(z) \varpi_0^{\text{(inv)}}(z) \alpha'(z) (11 z+13)}{\alpha(z)^2} \\[2pt]
G_{17}'(z) &= -\frac{2 G_{1}(z) \alpha(z) \varpi_0^{\text{(shift)}}(z) \left(35 z^3+103 z^2-16 z-2\right)}{5 \pi_0(z)z (z-1)  (9 z-1) (25 z-1)} \\
  &\quad - \frac{G_{10}(z) G_{12}(z)}{\big(\varpi_0^{\text{(inv)}}(z)\big)^2 z(z-9) (z-1) } + \frac{G_{11}(z) G_{3}(z)}{\alpha(z) z} \\
  &\quad - \frac{3 G_{11}(z) \left(525 z^4+1712 z^3+378 z^2-56 z+1\right)}{2 z(z-1)  (7 z+1) (9 z-1) (25 z-1)} \\
  &\quad + \frac{2 G_{3}(z) \pi_0(z) \varpi_0^{\text{(shift)}}(z) \left(35 z^3-28 z^2+15 z+4\right)}{5 \alpha(z) z^3} \\
  &\quad + \frac{4 G_{3}(z) \varpi_0^{\text{(shift)}}(z) \pi_0'(z) \left(35 z^3+103 z^2-16 z-2\right)}{5 \alpha(z) z^2} \\
  &\quad - \frac{2 G_{3}(z) \pi_0(z) \varpi_0^{\text{(shift)}}(z) \alpha'(z) \left(35 z^3+103 z^2-16 z-2\right)}{5 \alpha(z)^2 z^2} \\
  &\quad - \frac{q_{3}(z) \pi_0(z) \varpi_0^{\text{(shift)}}(z)}{10 z^3(z-1)^2  (9 z-1)^2 (25 z-1)^2} - \frac{q_{4}(z) \varpi_0^{\text{(shift)}}(z) \pi_0'(z)}{10z^2 (z-1)  (9 z-1) (25 z-1)} \\[2pt]
G_{18}'(z) &= -\frac{2 G_{1}(z) \alpha(z) \varpi_0^{\text{(inv)}}(z) z (11 z+13)}{\pi_0(z) (z-1) (9 z-1) (25 z-1)} - \frac{G_{10}(z) G_{11}(z) z}{\big(\varpi_0^{\text{(shift)}}(z)\big)^2 (z-1) (2 z-1) (7 z+1)} \\
  &\quad + \frac{2 G_{11}(z) \varpi_0^{\text{(inv)}}(z) z (7 z+5)}{\varpi_0^{\text{(shift)}}(z) (z-1) (2 z-1) (7 z+1)} + \frac{G_{12}(z) G_{3}(z)}{\alpha(z) z} \\
  &\quad - \frac{3 G_{12}(z) \left(525 z^4-2952 z^3-14 z^2-128 z+9\right)}{2 z(z-9) (z-1)  (9 z-1) (25 z-1)} - \frac{44 G_{3}(z) \pi_0(z) \varpi_0^{\text{(inv)}}(z)}{5 \alpha(z) z} \\
  &\quad + \frac{4 G_{3}(z) \varpi_0^{\text{(inv)}}(z) \pi_0'(z) (11 z+13)}{\alpha(z)} - \frac{2 G_{3}(z) \pi_0(z) \varpi_0^{\text{(inv)}}(z) \alpha'(z) (11 z+13)}{\alpha(z)^2} \\
  &\quad - \frac{q_{5}(z) \pi_0(z) \varpi_0^{\text{(inv)}}(z)}{10 z(z-1)^2  (9 z-1)^2 (25 z-1)^2} \\
  &\quad + \frac{\varpi_0^{\text{(inv)}}(z) \pi_0'(z) \left(39375 z^4-76940 z^3+18378 z^2-18732 z+1055\right)}{10 (z-1) (9 z-1) (25 z-1)} \\[2pt]
G_{19}'(z) &= \frac{3 \pi_0(z) \varpi_0(z) \left(37 z^2+10 z+1\right)}{5 z^2(z-1)^2 } - \frac{12 \varpi_0(z) \pi_0'(z) (13 z-1)}{5z (z-1) } \\[2pt]
G_{20}'(z) &= -\frac{G_{19}(z) z}{\varpi_0(z)^2 (z-1) (9 z-1)} \\[2pt]
G_{21}'(z) &= -\frac{G_{19}(z)}{\alpha(z) z} - \frac{12 \pi_0(z) \varpi_0(z) (13 z-1)}{5 \alpha(z) (z-1) z^2} \\[2pt]
G_{22}'(z) &=   \frac{G_{19}(z) G_{3}(z)}{z \alpha (z)}-\frac{\left(1703 z^3-141 z^2-27 z+1\right) G_{19}(z)}{8 (z-1) z^2 (9 z-1) (25 z-1)}-\frac{216 \varpi_0(z) \pi_0'(z)}{65 (z-1) z}\\
&\quad +\frac{12 (13 z-1) G_{3}(z) \pi_0(z) \varpi_0(z)}{5 (z-1) z^2 \alpha (z)}-\frac{3 q_6(z) \pi_0(z) \varpi_0(z)}{520 (z-1)^2 z^3 (9 z-1) (25 z-1)}\\[2pt]
G_{23}'(z) &= \frac{r_{1}(z) r_{4}(z) \varpi_0(z) (3 z-2)}{3z (z-1)  (z+1)^2} \\[2pt]
G_{24}'(z) &= \frac{G_{1}(z) \alpha(z) \varpi_0(z) \left(855 z^2-7 z-8\right)}{15 \pi_0(z) z(z-1)  (9 z-1) (25 z-1)} + \frac{12 G_{19}(z) G_{23}(z)^2 z}{\varpi_0(z)^2 (z-1) (9 z-1)} \\
  &\quad - \frac{2 G_{19}(z) G_{3}(z)}{\alpha(z) z} + \frac{G_{19}(z) \left(225 z^3-115 z^2+179 z-1\right)}{3 z (z-1) (z+1) (25 z-1)} \\
  &\quad - \frac{4 G_{23}(z)^2 G_{6}(z) z}{\varpi_0(z)^2 (z-1) (9 z-1)} + \frac{6 G_{23}(z)^2 \pi_0(z) \left(2829 z^2-618 z+29\right)}{5 \varpi_0(z) (z-1)^2 (9 z-1)^2} \\
  &\quad + \frac{2 G_{23}(z)^2 \pi_0(z) \varpi_0'(z) \left(855 z^2-7 z-8\right)}{5 \varpi_0(z)^2 (z-1) (9 z-1)} \\
  &\quad - \frac{2 G_{23}(z)^2 \pi_0'(z) \left(855 z^2-7 z-8\right)}{5 \varpi_0(z) (z-1) (9 z-1)} \\
  &\quad + \frac{8 G_{23}(z) G_{6}(z) r_{1}(z) r_{4}(z) (3 z-2)}{\varpi_0(z) (z-1) (z+1) (9 z-1)} \\
  &\quad - \frac{16 G_{23}(z) G_{7}(z) r_{1}(z) r_{4}(z) \varpi_0(z)}{z^2 (z+1)} - \frac{8 G_{23}(z) G_{7}(z) r_{1}(z) r_{4}(z) \varpi_0'(z) (3 z-2)}{z (z+1)} \\
  &\quad - \frac{4 G_{23}(z) r_{1}(z) r_{4}(z) \pi_0(z) \left(243 z^4-1008 z^3+47 z^2+150 z-8\right)}{15 z^2(z-1)^2  (z+1) (9 z-1)^2} \\
  &\quad + \frac{4 G_{23}(z) r_{1}(z) r_{4}(z) \pi_0'(z) \left(405 z^3+27 z^2-16 z-8\right)}{15 z(z-1)  (z+1) (9 z-1)} + \frac{2 G_{3}(z) G_{6}(z)}{3 \alpha(z) z} \\
  &\quad + \frac{G_{3}(z) \pi_0(z) \varpi_0(z) \left(-8289 z^3+1490 z^2+95 z-16\right)}{15 \alpha(z) z^3(z-1)  (9 z-1)} \\
  &\quad + \frac{G_{3}(z) \pi_0(z) \varpi_0'(z) \left(-855 z^2+7 z+8\right)}{15 \alpha(z) z^2} + \frac{G_{3}(z) \varpi_0(z) \pi_0'(z) \left(-855 z^2+7 z+8\right)}{15 \alpha(z) z^2} \\
  &\quad + \frac{G_{3}(z) \pi_0(z) \varpi_0(z) \alpha'(z) \left(855 z^2-7 z-8\right)}{15 \alpha(z)^2 z^2} \\
  &\quad + \frac{G_{6}(z) \left(225 z^3-1703 z^2+643 z-29\right)}{9z (z-1)  (z+1) (25 z-1)} - \frac{2 G_{7}(z) \varpi_0(z)^2 \left(33 z^2-42 z+5\right)}{3 z^3 (z+1)} \\
  &\quad - \frac{2 G_{7}(z) \varpi_0(z) \varpi_0'(z) (9 z-1) \left(3 z^2-12 z+5\right)}{3 z^2 (z+1)} \\
  &\quad + \frac{q_{7}(z) \pi_0(z) \varpi_0(z)}{90z^3 (z-1)^2  (z+1) (9 z-1)^2 (25 z-1)^2} \\
  &\quad + \frac{\pi_0(z) \varpi_0'(z) \left(855 z^2-7 z-8\right) \left(225 z^3-115 z^2+179 z-1\right)}{90z^2 (z-1)  (z+1) (25 z-1)} \\
  &\quad - \frac{q_{8}(z) \varpi_0(z) \pi_0'(z)}{90 z^2(z-1)  (z+1) (9 z-1) (25 z-1)} \\[2pt]
G_{25}'(z) &= \frac{2 G_{19}(z) G_{23}(z) z}{\varpi_0(z)^2 (z-1) (9 z-1)} - \frac{2 G_{23}(z) G_{6}(z) z}{3 \varpi_0(z)^2 (z-1) (9 z-1)} \\
  &\quad + \frac{G_{23}(z) \pi_0(z) \left(2829 z^2-618 z+29\right)}{5 \varpi_0(z) (z-1)^2 (9 z-1)^2} + \frac{G_{23}(z) \pi_0(z) \varpi_0'(z) \left(855 z^2-7 z-8\right)}{15 \varpi_0(z)^2 (z-1) (9 z-1)} \\
  &\quad + \frac{G_{23}(z) \pi_0'(z) \left(-855 z^2+7 z+8\right)}{15 \varpi_0(z) (z-1) (9 z-1)} + \frac{2 G_{6}(z) r_{1}(z) r_{4}(z) (3 z-2)}{3 \varpi_0(z) (z-1) (z+1) (9 z-1)} \\
  &\quad - \frac{4 G_{7}(z) r_{1}(z) r_{4}(z) \varpi_0(z)}{3 z^2 (z+1)} - \frac{2 G_{7}(z) r_{1}(z) r_{4}(z) \varpi_0'(z) (3 z-2)}{3 z (z+1)} \\
  &\quad + \frac{r_{1}(z) r_{4}(z) \pi_0(z) \left(-243 z^4+1008 z^3-47 z^2-150 z+8\right)}{45z^2 (z-1)^2  (z+1) (9 z-1)^2} \\
  &\quad + \frac{r_{1}(z) r_{4}(z) \pi_0'(z) \left(405 z^3+27 z^2-16 z-8\right)}{45 z(z-1)  (z+1) (9 z-1)} \\[2pt]
G_{26}'(z) &= \frac{6 G_{1}(z) \alpha(z) \varpi_0(z) (15 z-7)}{5 \pi_0(z) (z-1) (9 z-1) (25 z-1)} - \frac{4 G_{3}(z) G_{6}(z)}{\alpha(z) z} \\
  &\quad - \frac{2 G_{3}(z) \pi_0(z) \varpi_0(z) \left(387 z^2-230 z+35\right)}{5 \alpha(z)z^2 (z-1)  (9 z-1)} \\
  &\quad - \frac{6 G_{3}(z) \pi_0(z) \varpi_0'(z) (15 z-7)}{5 \alpha(z) z} - \frac{6 G_{3}(z) \varpi_0(z) \pi_0'(z) (15 z-7)}{5 \alpha(z) z} \\
  &\quad + \frac{6 G_{3}(z) \pi_0(z) \varpi_0(z) \alpha'(z) (15 z-7)}{5 \alpha(z)^2 z} \\
  &\quad + \frac{4 G_{6}(z) \left(1023 z^2-62 z-1\right)}{3z (z-1)  (9 z-1) (25 z-1)} \\
  &\quad + \frac{8 G_{7}(z) \varpi_0(z)^2 (3 z+2)}{3 z^3} - \frac{16 G_{7}(z) \varpi_0(z) \varpi_0'(z) (3 z+1)}{3 z^2} \\
  &\quad + \frac{q_{9}(z) \pi_0(z) \varpi_0(z)}{15z^3 (z-1)^2  (9 z-1) (25 z-1)^2} \\
  &\quad + \frac{6 \pi_0(z) \varpi_0'(z) (15 z-7) \left(241 z^2-50 z+1\right)}{5z (z-1)  (9 z-1) (25 z-1)} \\
  &\quad + \frac{\varpi_0(z) \pi_0'(z) \left(-14985 z^4+8270 z^3-2692 z^2+194 z-3\right)}{5 z^2 (z-1) (9 z-1) (25 z-1)} \\[2pt]
G_{27}'(z) &= \frac{G_{19}(z)}{8 z} + \frac{\pi_0(z) \varpi_0(z) \left(733 z^2+23 z-20\right)}{240z^3 (z-1) } + \frac{\pi_0(z) \varpi_0'(z) \left(855 z^2-7 z-8\right)}{240 z^2} \\
  &\quad + \frac{\varpi_0(z) \pi_0'(z) \left(-495 z^3+1442 z^2-191 z+12\right)}{240 z^2(z-1) } \, ,
\end{align*}
where we abbreviated
% Polynomials q_k(z) abbreviated above
\begin{align*}
q_{1}(z) &= 2562890625 z^{12} - 9146671875 z^{11} + 29858270625 z^{10} - 4836275325 z^9 \\
  &\quad - 3663006612 z^8 + 1595702658 z^7 - 43328414 z^6 - 25045562 z^5 + 5506553 z^4 \\
  &\quad - 624143 z^3 + 36365 z^2 - 953 z + 10 \, , \\[2pt]
q_{2}(z) &= 50625 z^6 + 302850 z^5 + 77427 z^4 + 12044 z^3 - 569 z^2 - 14 z + 5\, , \\[2pt]
q_{3}(z) &= 3543750 z^9 - 10993500 z^8 + 31721160 z^7 - 8994033 z^6 + 5964198 z^5 - 4310691 z^4 \\
  &\quad + 826036 z^3 - 64215 z^2 + 2040 z - 25\, , \\[2pt]
q_{4}(z) &= 31500 z^6 - 45330 z^5 + 78471 z^4 - 16520 z^3 - 12690 z^2 + 1458 z - 25\,, \\[2pt]
q_{5}(z) &= 7707375 z^6 + 15401250 z^5 - 8747659 z^4 + 3738668 z^3 - 423063 z^2 + 18322 z - 173\,, \\[2pt]
q_{6}(z) &= 283275 z^5+609197 z^4-32994 z^3-13206 z^2+1639 z-39\,, \\[2pt]
q_{7}(z) &= 608623875 z^8 - 218761425 z^7 + 425191833 z^6 - 95774533 z^5 + 1393635 z^4 \\
  &\quad + 1397533 z^3 - 129821 z^2 + 3513 z - 34\,, \\[2pt]
q_{8}(z) &= 1002375 z^6 - 911925 z^5 + 1181112 z^4 - 505706 z^3 + 66807 z^2 - 3249 z + 26\,, \\[2pt]
q_{9}(z) &= 904500 z^6 + 1253475 z^5 - 439403 z^4 + 43230 z^3 + 8190 z^2 - 529 z + 9\,.
\end{align*}

\newpage

%---------bibliography if using bibtex
\bibliographystyle{JHEP} % doi-Link, when doi  = {10.1039/C3CC46767H} in .bib file
\bibliography{biblio} % with XeLaTex: needs the same name as main texfile

\end{document}